\documentclass[twocolumn]{aastex62}
\newcommand{\msun}{M_{\odot}}
\newcommand{\lsun}{L_{\odot}}

\newcommand{\kms}{km~s$^{-1}$}

\newcommand{\hi}{H{\sc\,i}}

\graphicspath{{./}{figures/}}

\received{January 20, 2021}
\accepted{September 24, 2021}

\shorttitle{The ALFALFA Almost Dark Galaxy AGC~229101}
\shortauthors{Leisman et al.}

\begin{document}

\title{The ALFALFA Almost-Dark Galaxy AGC~229101: A Two Billion Solar Mass HI Cloud\\
with a Very Low Surface Brightness Optical Counterpart}

\correspondingauthor{Lukas Leisman}
\email{luke.leisman@valpo.edu}

\author[0000-0001-8849-7987]{Lukas Leisman}
\affiliation{Department of Physics and Astronomy, Valparaiso University,  1610 Campus Drive East, Valparaiso, IN 46383, USA}
\affiliation{Department of Astronomy, University of Illinois, 1002 W. Green St., Urbana, IL 61801, USA}

\author[0000-0001-8283-4591]{Katherine L. Rhode}
\affiliation{Department of Astronomy, Indiana University, 727 East
 Third Street, Bloomington, IN 47405, USA} 

\author[0000-0002-1895-0528]{Catherine Ball}
\affiliation{Department of Physics \& Astronomy, Macalester College, 1600 Grand Avenue, Saint Paul, MN 55105}
\affiliation{Cornell Center for Astrophysics and Planetary Science, 
Space Sciences Building, Cornell University, Ithaca, NY 14853, USA}

\author[0000-0002-0786-7307]{Hannah J. Pagel}
\affil{Department of Astronomy, Indiana University, 727 East
 Third Street, Bloomington, IN 47405, USA}

\author[0000-0002-1821-7019]{John M. Cannon}
\affiliation{Department of Physics \& Astronomy, Macalester College, 1600 Grand Avenue, Saint Paul, MN 55105}

\author[0000-0001-8483-603X]{John J. Salzer}
\affiliation{Department of Astronomy, Indiana University, 727 East
 Third Street, Bloomington, IN 47405, USA}

\author[0000-0001-9165-8905]{Steven Janowiecki}
\affiliation{University of Texas, Hobby-Eberly Telescope, McDonald Observatory, TX 79734, USA}

\author[0000-0003-4364-0799]{William F. Janesh}
\affiliation{Department of Astronomy, Case Western Reserve University, 10900 Euclid Avenue, Cleveland, OH 44106, USA}

\author{Gyula I. G. J\'ozsa}
\affil{South African Radio Astronomy Observatory, 2 Fir Street, Black River Park, Observatory, Cape Town, 7925, South Africa}
\affil{Department of Physics and Electronics, Rhodes University, PO Box 94, Makhanda, 6140, South Africa}

\author{Riccardo Giovanelli}
\affiliation{Cornell Center for Astrophysics and Planetary Science, 
Space Sciences Building, Cornell University, Ithaca, NY 14853, USA}

\author[0000-0001-5334-5166]{Martha P. Haynes}
\affiliation{Cornell Center for Astrophysics and Planetary Science, 
Space Sciences Building, Cornell University, Ithaca, NY 14853, USA}

\author[0000-0002-9798-5111]{Elizabeth A. K. Adams}
\affiliation{ASTRON, Netherlands Institute for Radio Astronomy, Oude Hoogeveensedijk 4, 7991 PD Dwingeloo, The Netherlands}
\affiliation{Kapteyn Astronomical Institute, University of Groningen, Landleven 12, 9747 AD, Groningen, The Netherlands}

\author[0000-0001-6389-5639]{Laurin Gray}
\affiliation{Department of Astronomy, Indiana University, 727 East
 Third Street, Bloomington, IN 47405, USA}

\author[0000-0002-3222-2949]{Nicholas J. Smith}
\affiliation{Department of Astronomy, Indiana University, 727 East
 Third Street, Bloomington, IN 47405, USA}

\begin{abstract}
We present results from deep \hi\ and optical imaging of AGC~229101, an unusual \hi\ source detected at v$_{\rm helio}$ $=$ 7116 \kms\ in the ALFALFA survey. Initially classified as a candidate ``dark'' source because it lacks a clear optical counterpart in SDSS or DSS2 imaging, AGC~229101 has $10^{9.31\pm0.05}\msun$ of \hi, but an \hi\ line width of only 43$\pm$9~\kms. Low resolution WSRT imaging and higher resolution VLA B-array imaging show that the source is significantly elongated, stretching over a projected length of $\sim$80~kpc. The HI imaging resolves the source into two parts of roughly equal mass. WIYN pODI optical imaging reveals a faint, blue optical counterpart coincident with the northern portion of the \hi. The peak surface brightness of the optical source is only $\mu_{g}$ $\sim$ 26.6 mag~arcsec$^{-2}$, well below the typical cutoff that defines the isophotal edge of a galaxy, and its estimated stellar mass is only
$10^{7.32\pm0.33}\msun$, 
yielding an overall neutral gas-to-stellar mass ratio of M$_{\rm HI}$/M$_*=$~98$_{+111}\atop^{-52}$.
We demonstrate the extreme nature of this object by comparing its properties to those of other \hi-rich sources in ALFALFA and the literature.  We also explore potential scenarios that might  explain the existence of AGC~229101, including a tidal encounter with neighboring objects and a merger of two dark \hi\ clouds.
\end{abstract}



\section{Introduction} \label{sec:intro}

The Arecibo Legacy Fast ALFA blind \hi\ survey (ALFALFA; \citealp{giovanelli05a}, \citealp{haynes18a}) has detected and cataloged $\sim$31,500 extragalactic \hi\ sources out to a redshift of $z\sim 0.06$, providing a robust characterization of the \hi\ properties of galaxies in the nearby universe. One of the many scientific motivations for ALFALFA was to investigate the possible existence of gas-rich but nearly starless ``dark" galaxies (e.g., \citealp{giovanelli05a,giovanelli10a}). Scaling relations between the \hi\ and stellar contents of galaxies suggest that there are strong ties between atomic gas content and star formation laws (e.g., \citealp{huang12b}), and theoretical predictions are mixed as to the potential existence of sources that are \hi-rich while also being optically dark \citep{verde02a,taylor05a,crain17a}. 

Only $\sim$1\% of ALFALFA sources are not readily associated with optical counterparts in existing catalogs or surveys like the Sloan Digital Sky Survey (SDSS; \citealp{eisenstein11a}).  The majority of these sources can clearly be identified as either debris from tidal interactions between galaxies, or as OH megamasers masquerading as \hi\
\citep{leismanthesis, suess16a}. The remaining objects fall into two main categories.  The first is Ultra-Compact High-Velocity Clouds (e.g., \citealp{giovanelli10a,adams13a,janesh19a}), which have velocities and other properties (e.g., HI sizes, masses) that make them good candidates for as-yet unidentified gas-rich dwarf galaxies located in and around the Local Group. The second category is ``Almost Dark galaxies" (e.g., \citealp{cannon15a,janowiecki15a}), which also lack an unambiguous optical counterpart (although some do show faint optical emission nearby that may or may not be associated with the \hi\ source) but have a much wider range of masses and distances than the UCHVCs.  Nearly all of the Almost Dark sources have ultimately been associated with very low surface brightness optical counterparts, some of which have similar properties and stellar populations to so-called ``ultra-diffuse galaxies" (UDGs; see \citealp{vandokkum15a}, \citealp{leisman17a} and Section~\ref{sec:discussion:UDG} of this paper), and others which may be tidal in nature \citep{lee-waddell16a,leisman16a}.  

A few of the Almost Dark objects have been particularly difficult to classify and interpret, and have required dedicated follow-up observations in order to determine their nature and properties.  For example, the object Coma~P (AGC~229385) was detected by ALFALFA with a heliocentric recession velocity of 1348~\kms\ and a signal-to-noise ratio (S/N) of 99, and appeared as an ultra-low surface brightness object (peak surface brightness $\mu_g$ $=$ 26.4 mag arcsec$^{-2}$) in deep follow-up imaging with the WIYN 3.5-m Observatory \citep{janowiecki15a}.  The object was later resolved into stars in Hubble Space Telescope (HST) imaging presented in \cite{brunker19a}, revealing an unexpectedly small distance (5.5~Mpc) and large peculiar velocity \citep{ball18a,anand18a,brunker19a}. Additionally, \cite{ball18a} presented resolved \hi\ imaging, which revealed a \hi\ diameter of $\sim$4~kpc, large for both its \hi\ and stellar mass, and complicated gas dynamics. Thus, Coma~P has proved difficult to understand both in terms of its distance, stellar properties, \hi\ properties, and its large gas mass to optical light ratio ($\sim$30).

In this paper, we present results from observations of an ALFALFA Almost Dark source that 
also possesses unusual properties and has likewise proved difficult to classify and explain. Like Coma~P, AGC~229101 is detected strongly in the ALFALFA survey, and has no readily identifiable optical counterpart in corresponding optical imaging from surveys like SDSS. 
Unlike Coma~P, however, AGC~229101 has a recession velocity of 7116~\kms, which is large enough to provide a reasonably accurate Hubble flow distance even if its peculiar velocity is several hundred \kms. This allows us to derive fundamental quantities that are distance-dependent, such as HI mass, stellar mass, absolute magnitude, and physical size for the object. The distance to AGC~229101 in the ALFALFA Extragalactic \hi\ Source Catalog is 105.9 $\pm$ 2.2~Mpc; this is calculated by combining the recessional velocity in the CMB reference frame with an H$_0$ value of 70~km~s$^{-1}$~Mpc$^{-1}$ \citep{haynes18a}.  Follow-up optical observations with the WIYN 3.5-m telescope\footnote{The WIYN Observatory is a joint facility of the NSF’s National Optical-Infrared Astronomy Research Laboratory, Indiana University,
the University of Wisconsin-Madison, Pennsylvania State University,
the University of Missouri, the University of California-Irvine and
Purdue University.}
and \hi\ synthesis imaging reveal an even lower surface brightness optical counterpart and more extended gas distribution, and a larger \hi\ gas mass to optical light ratio, than those of Coma~P.  In this paper, we present results from these follow-up observations, which paint a picture of an object that is extreme in its properties when compared to the 
over 31,000 extragalactic ALFALFA detections. We also explore potential scenarios that may help us understand AGC~229101 and its origins. 

The paper is organized as follows. In Section \ref{sec:observations} we discuss the \hi\ and optical observations of AGC~229101.  In Section \ref{sec:results}, we present the results of the observations, including the \hi\ gas content and stellar properties of AGC~229101, and then examine the galaxy environment surrounding the object. In the final section of the paper, we compare AGC~229101 to other sources, present potential interpretations of our observations, and speculate about the nature and origin of this extreme source.  Throughout the paper, we assume a $\Lambda$CDM cosmological model, with $H_0 =$ 70~\kms, $\Omega_M=0.3$, and $\Omega_{\Lambda}=0.7$, and the \citep{haynes18a} distance to AGC~229101 of 105.9$\pm$2.2~Mpc. 

\section{Observations and Analysis}
\label{sec:observations}

\begin{figure*}[t!]
\centering
\includegraphics[width=0.8\textwidth]{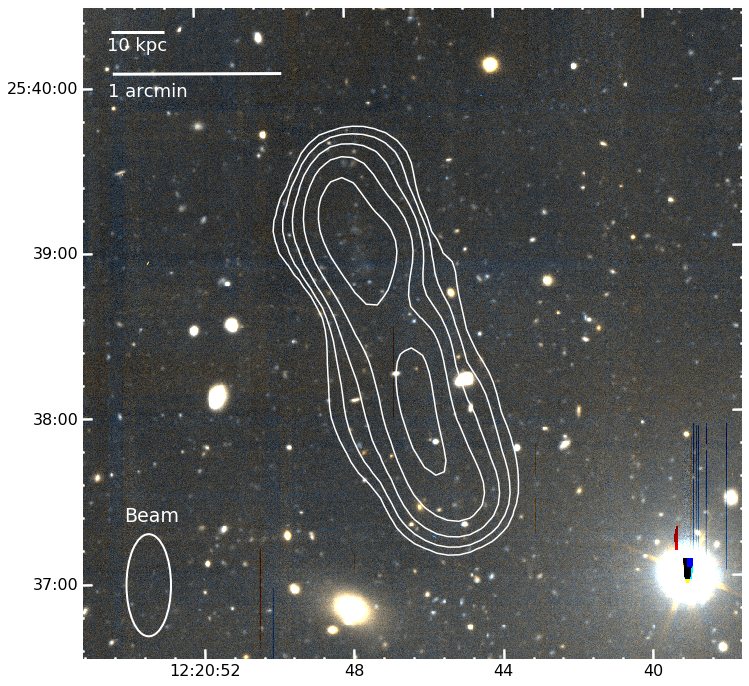}
\vspace{-0.1cm}
\caption{WIYN 3.5-m pODI combined g- and r-band color image of AGC~229101 with \hi\ column density
contours from WSRT only imaging at 1, 2, 4, 8, and 16 $\times$10$^{19}$~cm$^{-2}$ overlaid in white. 
The image is oriented N-up, E-left.  Faint, diffuse optical emission is barely visible at the location of the northern \hi\ column density peak, while the \hi\ emission stretches over $\sim$80~kpc. The early-type galaxy to the southeast of AGC~229101 is unrelated and has an SDSS redshift of 0.176.
\label{fig:main}
}
\end{figure*}

\subsection{HI Data}
\label{sec:data.hi}

The ALFALFA detection of AGC~229101 revealed a source at RA, Dec $=$ 12h20m48.7s, +25d38$\arcmin$19$\arcsec$ with a signal-to-noise ratio (S/N) of 13.2 and a heliocentric recessional velocity of 7116$\pm$4~\kms. The source was immediately noted for its high \hi\ mass and the lack of an optical counterpart in existing survey data. The ALFALFA data yield a log($M_{\rm HI}/M_{\odot}$) value of 9.31, comparable with that of the Milky Way, but with a narrow \hi\ line width for its mass ($43\pm9$~\kms), and no easily identifiable optical counterpart in SDSS or Digitized Sky Survey 2 (DSS2) imaging.  

Due to its apparently extreme properties, AGC~229101 was included in a set of exploratory observations of selected ALFALFA Almost Dark sources carried out with the Westerbork Synthesis Radio Telescope (WSRT) (program R13B/001; PI Adams). AGC~229101 was observed in 12-hour pointings in two polarizations, with a 1024-channel 10~MHz bandpass centered on the central \hi\ velocity measured in ALFALFA. 

The data reduction process for AGC~229101 and other sources in the exploratory WSRT sample is described in detail elsewhere (\citealp{janowiecki15a}; \citealp{leisman16a}, \citealp{leisman17a}). The reduction was accomplished with an automated pipeline based on the MIRIAD \citep{sault95a} data software (see \citealp{serra12a}, \citealp{wang13a}). The pipeline includes automatic RFI flagging, primary bandpass calibration, and iterative deconvolution of the data with the CLEAN algorithm to apply a self-calibration. This process produces cubes with three different robustness weightings, r=0.0, r=0.4, and r=6.0, and bins the data to a velocity resolution of 6.0~\kms\ after Hanning smoothing.

After initial reduction of the WSRT data indicated that AGC~229101 has a large extent, the source was included in a high resolution follow-up program (15A-307; P.I. Leisman) carried out with the Karl G. Jansky Very Large Array (VLA) in 2015. We observed the source for three five-hour observing blocks in the B configuration, using the WIDAR correlator in dual polarization mode with a single 8-MHz-wide sub-band with 1024 channels.  This yielded a native channel width of 1.7~\kms. 

The VLA data reduction follows the same procedure detailed in \citet{ball18a}. We use standard procedures in the CASA (Common Astronomy Software Applications; \citealp{mcmullin07a}) package, including flagging of the visibilities, calibration, and continuum subtraction. We imaged the calibrated uv data using the CLEAN task in CASA, with a Briggs robust weighting of 2.0 to maximize sensitivity. We also imaged the uv data applying an 8 k$\lambda$ taper to create an lower resolution image using the shorter uv spacings. Importantly, in both cases we follow the procedure in \citet{ball18a} to image together the uv data set from WSRT with the VLA data set, creating maps that includes baselines with a wide range of spacings, increasing the sensitivity of the VLA observations to extended emission. We also improved the localization of the extended flux by using the multiscale clean option \citep{cornwell08a}. The resulting cleaned combined VLA and WSRT images have synthesized beams of 5.1\arcsec\ and 13.1\arcsec\ with and without the 8~k$\lambda$ taper respectively.

We create \hi\ total intensity maps from all cleaned images by masking the image cubes at 2$\sigma$ and then summing along the velocity axis. We then blank the noise in the moment 0 maps using a 2$\sigma$ mask created with smoothed versions of the moment 0 maps.
We convert these maps to \hi\ column densities assuming optically thin \hi\ gas that fills the beam, 
and also produce \hi\ moment one maps (representing velocity fields) from the masked cubes. The resulting \hi\ images and velocity maps are discussed in Section \ref{sec:results.hi}. 

\subsection{Optical Data}
\label{sec:data.optical}
Observations of AGC~229101 were obtained on 1 April 2014 with the partially-populated One Degree Imager (pODI; \citealp{harbeck14a}) on the 3.5-meter WIYN telescope at Kitt Peak National Observatory. The pODI camera (which has since been upgraded with a larger detector array) included nine orthogonal transfer arrays (OTAs) laid out in a 3 x 3 grid in the center of the focal plane. Four additional OTAs were positioned outside the central grid and used for imaging guide stars. Each individual OTA consists of an 8 x 8 arrangement of CCD detectors.  The pixel scale of the CCDs was 0.11\arcsec~pixel$^{^-1}$ and the field-of-view of the central 3 x 3 grid of OTAs was approximately 24\arcmin\ x 24\arcmin. We observed AGC~229101 in the SDSS \textit{g} and \textit{r} filters. In order to fill in the gaps between the CCD detectors, we divided the observations into nine 300-second exposures and dithered the telescope between exposures.  The total integration time was 45 minutes per filter.

Our images were reduced using the QuickReduce data reduction pipeline \citep{kotulla14a} within the One Degree Imager Pipeline, Portal, and Archive (ODI-PPA) system\footnote{The ODI Portal, Pipeline, and Archive (ODI-PPA) system is a joint development project of the WIYN Consortium, Inc., in partnership with Indiana University's Pervasive Technology Institute (PTI) and NSF’s NOIRLab.} \citep{gopu14a}. The QuickReduce pipeline performs the following reduction tasks: masks saturated pixels; corrects for crosstalk and persistence; subtracts the overscan signal; corrects for nonlinearity; applies the bias, dark, and flat-field corrections; applies a pupil ghost correction; and removes cosmic rays. After the QuickReduce step, an illumination correction was applied to the images, and the images were then background-subtracted, scaled to a common flux level, and then combined, and the appropriate sky background was restored.  The final result was a single science-ready image in each filter. The average full-width at half-maximum (FWHM) of the point spread function (PSF) is 0.8\arcsec\ in the final combined \textit{g} image and 0.9\arcsec\ in the final combined \textit{r} image. Photometric measurements of SDSS stars that appeared within the pODI images were used to derive calibration coefficients (zero points and color terms) that were later applied to our photometric measurements of the other sources in the frames.

The main results of our follow-up imaging campaign are illustrated in Figure~\ref{fig:main}, which shows a color image constructed from the \textit{g} and \textit{r} WIYN pODI images of AGC~229101, with low resolution \hi\ column density contours from WSRT observations shown as white solid lines overlaid on the image. The column density levels shown in the figure correspond to 1, 2, 4, 8, and 16 x 10$^{19}$~cm$^{-2}$.  
Our deep \hi\ WSRT and VLA observations reveal that the \hi\ distribution of AGC~229101 is double-peaked and extends $\sim$80~kpc in roughly the north-south direction. Extremely faint, blue, 
low surface brightness optical emission that coincides with the northern peak of the \hi\ column density distribution is barely visible in the color image.  Details of the analysis and measurements we have derived from the \hi\ and optical data are given in the next section.

\section{Results}
\label{sec:results}

\subsection{The HI content of AGC~229101}
\label{sec:results.hi}

\begin{figure*}[t!]
\centering
\includegraphics[width=0.455\textwidth]{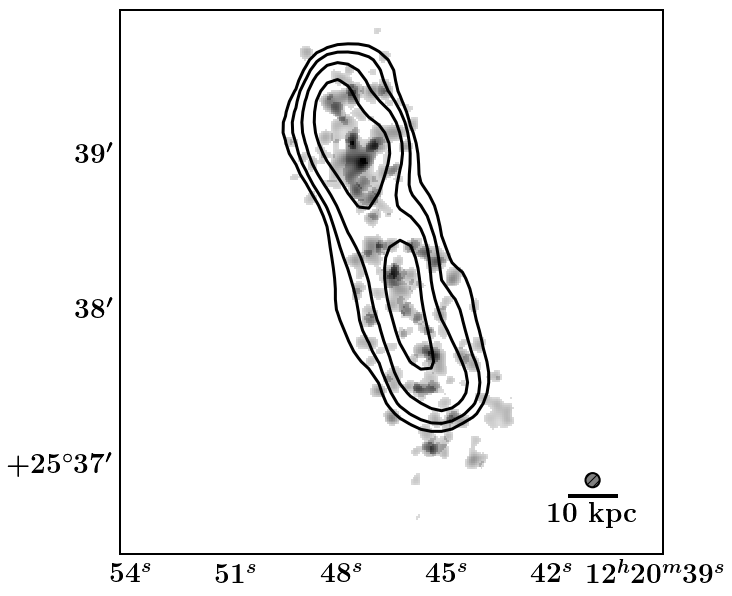}
\includegraphics[width=0.47\textwidth]{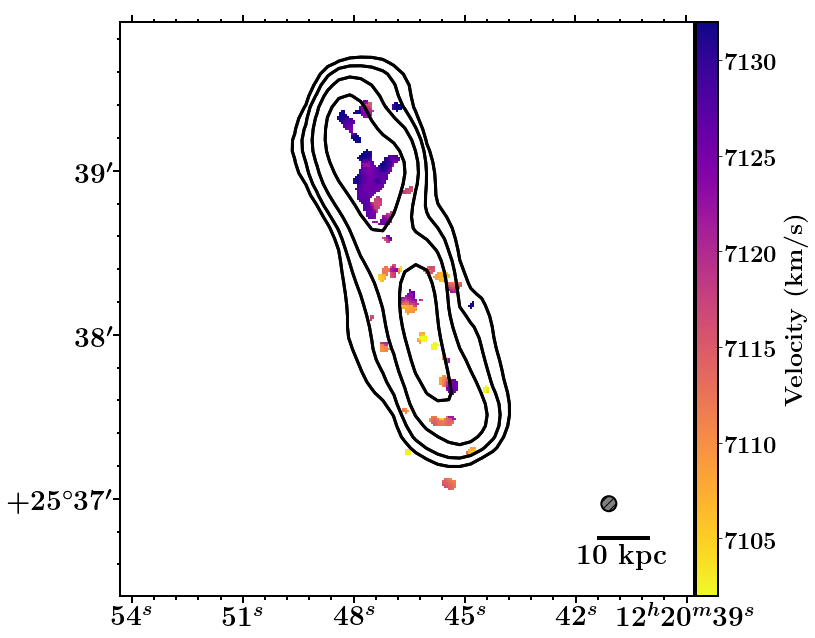}
\includegraphics[width=0.455\textwidth]{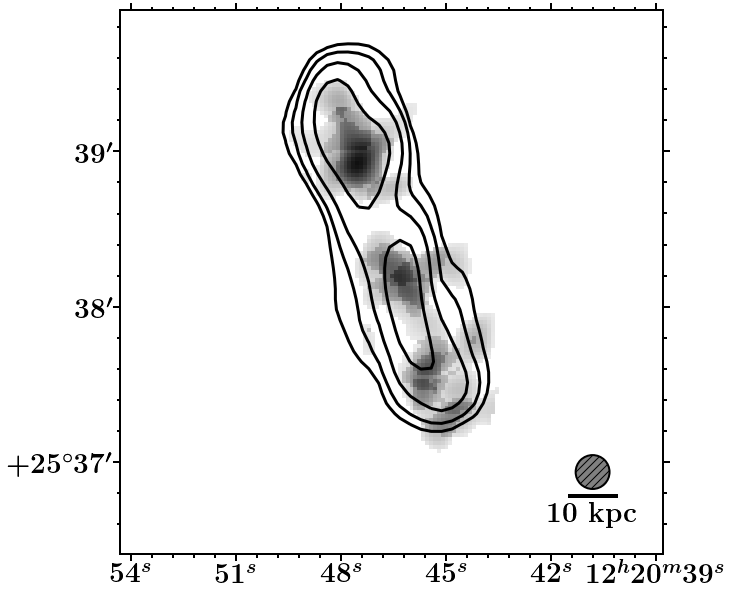}
\includegraphics[width=0.47\textwidth]{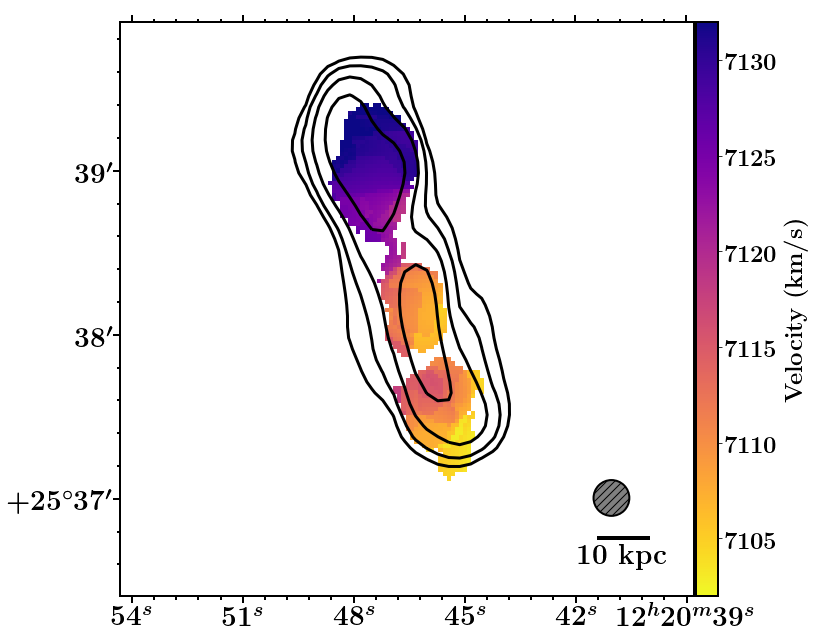}
\includegraphics[width=0.42\textwidth]{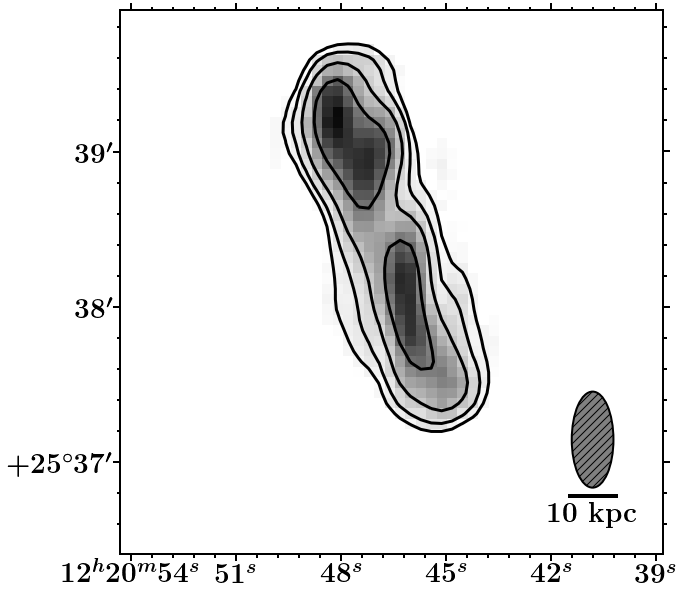} \hspace{0.2cm}
\includegraphics[width=0.47\textwidth]{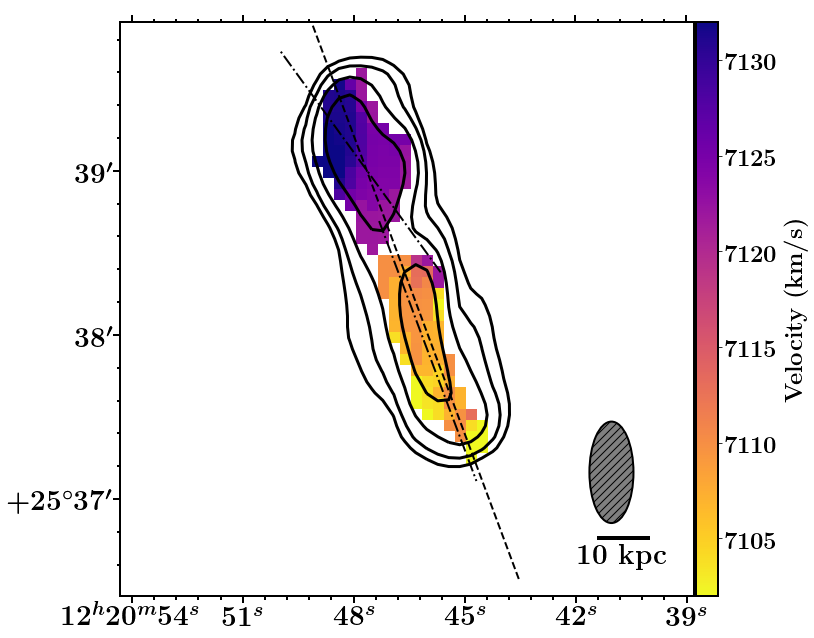}

\vspace{-0.1cm}
\caption{Left panels: \hi\ moment zero column density maps at high- (top: 5.5\arcsec\ beam), mid- (center: 13.1\arcsec\ beam) and low- resolutions (bottom: 16$\times$37\arcsec\ beam). The high and mid resolution maps are from combined VLA B-array and WSRT imaging, and the low resolution map is from WSRT only, Briggs r=0.4 imaging. Beam sizes are shown with hashed gray circles in the lower right on each plot. The column density scale in the top image is 0.5-4.5$\times$10$^{20}$atoms~cm$^{-2}$, and is 0.15-3.2$\times$10$^{20}$atoms~cm$^{-2}$ in the center image. Column density contours spaced in powers of 2 from 0.2-1.6$\times$10$^{20}$atoms~cm$^{-2}$ from the WSRT moment 0 map are shown on all plots for direct comparison. Moment maps are masked to only include emission detected above 2$\sigma$, as discussed in the text. Right panels: Corresponding moment one velocity maps from combined VLA and WSRT (top and center), and WSRT only (bottom), imaging. 
Resolutions correspond to those of the moment zero maps to the left, masked at the 5$\sigma$ level to highlight the motions of the highest signal-to-noise gas. The dashed line in the lower right panel shows the location of the position velocity slice shown in Figure \ref{fig:pv}, and the dot dashed lines show the locations of the northern and southern PV slices in Figure \ref{fig:pv}. 
\label{fig:mmaps}
}
\end{figure*}

\begin{figure*}[t!]
\centering
\includegraphics[width=0.33\textwidth]{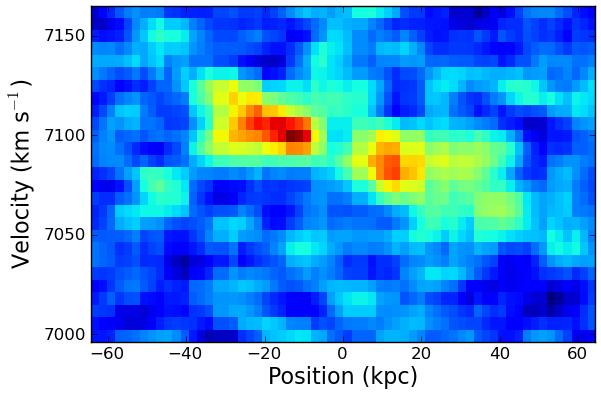}
\includegraphics[width=0.32\textwidth]{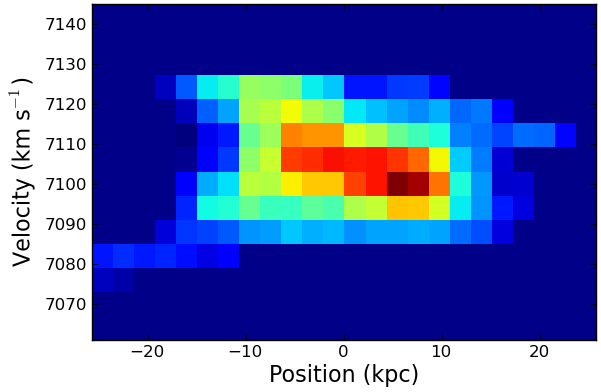}
\includegraphics[width=0.32\textwidth]{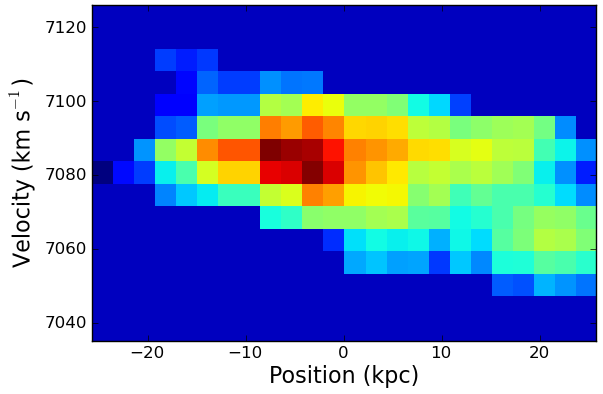}
\vspace{-0.1cm}
\caption{Position velocity slices made along the major axis of the WSRT (lowest resolution) data. Slices for each are 10\arcsec\ wide. The left panel shows the position versus velocity plot across the full extent of the source; the central panel shows just the northern portion of the source, and the right hand panel shows just the southern portion. 
\label{fig:pv}
}
\end{figure*}

Figure \ref{fig:mmaps} shows the resolved \hi\ total flux maps from WSRT and combined VLA and WSRT imaging at three different resolutions. All of the maps show a very elongated \hi\ structure, stretching $\sim$80~kpc in roughly the north-south direction. 
%
The emission is resolved into two primary overdensities that appear connected in the low resolution imaging (which is sensitive to low column densities) but that appear as two separate clumps at higher resolution. At the highest resolution (5.5\arcsec = 2.6 kpc beam), the gas is resolved into a large number of individual clumps and clouds, though this may primarily be the effect of the low sensitivity of the image, since much of the emission is only detected at the 2-3$\sigma$ level in the cleaned cube at this resolution. 

We note that the high and medium resolution images recover 57\% and 65\% of the ALFALFA flux respectively, while the low resolution WSRT image recovers 83\% of the ALFALFA flux. 
This implies that there is low column density emission detected in the single dish image that is below the sensitivity of the interferometric image. Thus, the apparent separation of the northern and southern clouds in the higher resolution data may simply be the result of the more limited sensitivity of these observations. 

The northern portion of the \hi\ has a peak column density of 5.2$\times 10^{20}$~atoms~cm$^{-2}$ in the highest resolution B-array-only imaging,  while the southern portion has a peak column density of 4.2$\times 10^{20}$ atoms~cm$^{-2}$. The northern portion is somewhat less elongated, with a length of $\sim$35~kpc, and with a more symmetric \hi\ distribution. The southern portion extends $\sim$45~kpc along its major axis. The two portions have approximately equal \hi\ mass; selectively masking just the northern component results in a flux that is 49.8\% of the total WSRT flux, and, the southern component is 47.1\% of the total WSRT flux, with a small amount of low signal to noise flux between the two components. Applying these percentages to the total \hi\ mass calculated from the ALFALFA flux results in estimated component masses of $\log(M_{\rm HI}/M_{\odot}) = 9.01$ and 8.99 for the northern and southern clumps respectively.   

The two main components of the source appear to be at similar, but slightly offset, velocities ($\Delta V\sim26$ \kms). The panels on the right side of Figure \ref{fig:mmaps} show moment one velocity maps of the resolved \hi\ at three different resolutions. There is a marked but shallow velocity gradient across the entire source, as well as across both the northern and southern clumps (the line widths at 50\% of the peak flux are W$_{50}=34$ \kms\ and 41 \kms\ for the two clumps respectively).  We note that the apparent change in the direction of the velocity gradient of the northern component in the low resolution WSRT image versus the higher resolution images is likely a resolution effect, since the elongation of the WSRT beam is in the direction of the gradient seen in the higher resolution maps.
Figure \ref{fig:pv} shows position velocity slices through the major axis of the entire source, as well as the northern and southern clumps individually. Both clumps show a velocity gradient, though the gradient is somewhat narrow and on the order of the velocity dispersion in the beams.

\begin{deluxetable*}{lrlr}
\tablecaption{Measured and Derived Properties of AGC~229101}
\setlength{\tabcolsep}{24pt}
\tablehead{
\colhead{Quantity (units)} &
\colhead{Value} &
\colhead{Stellar Quantity (units)} &
\colhead{Value} 
}
\startdata
RA (h~m~s, J2000) 			& 12:20:46.8 & $\mu_{g,peak}$ (mag~arcsec$^{-2}$)	& 26.58 $\pm$ 0.03 \\
Dec ($^{\circ}$ $\arcmin$ $\arcsec$, J2000) & +25:38:24.4 & $\mu_{r,peak}$ (mag~arcsec$^{-2}$)	& 26.78 $\pm$ 0.06 \\ 
V$_{h,50}$ (km~s$^{-1}$)    & 7116$\pm$4    & m$_{g,0}$ (mag)	& 22.01 $\pm$ 0.18 \\ 
W$_{50}$ (km~$s^{-1}$) 	    & 43 $\pm$ 9    & m$_{r,0}$ (mag)	& 21.58 $\pm$ 0.19 \\ 
Flux (Jy-km~s$^{-1}$)  	    & 0.78 $\pm$ 0.05 & M$_{g,0}$ (mag)	& $-$13.11 $\pm$ 0.18 \\ 
Distance (Mpc)			    & 105.9 $\pm$ 2.2 & M$_{r,0}$ (mag) 	& $-$13.55 $\pm$ 0.19 \\ 
$\log_{10}(M_{\rm HI}/\msun)$ & 9.31 $\pm$ 0.05 & $(g-r)_{0}$ (mag; large aperture) & 0.44 $\pm$ 0.26\\ 
N$_{\rm HI,peak}$ ($10^{20}$ cm$^{-2}$) & 5.1 $\pm$ 1.1 &$(g-r)_0$ (mag; small aperture)& 0.06 $\pm$ 0.13 \\ 
R$_{\rm HI}$ (kpc)	 	& 33.5 $\pm$ 2.8 &  V (mag) 		& 21.74 $\pm$ 0.24 \\ 
R$_{\rm max}$ (kpc) 	& 39.6 $\pm$ 2.8 & B-V (mag; large aperture) & 0.65 $\pm$ 0.26 \\ 
M$_{\rm HI}$/M$_*$ 		& 98$_{+111}\atop^{-52}$ & B-V (mag; small aperture) & 0.28 $\pm$ 0.12 \\ 
M$_{\rm HI}$/L$_g$ ($\msun/\lsun$)		& 105$_{+20}\atop^{-16}$  & $\log_{10}(M_*/\msun)$ & 7.32 $\pm$ 0.33 \\ 
M$_{\rm HI}$/L$_r$ ($\msun/\lsun$)		& 107$_{+20}\atop^{-16}$ & $r_{h}$ (arcsec) & 5.9 $\pm$ 1.3\\ 
M$_{\rm HI}$/L$_B$ ($\msun/\lsun$)		& 110$_{+27}\atop^{-23}$ & $r_{h}$ (kpc) & 3.0 $\pm$ 0.7
\enddata
\tablecomments{
The table lists, in this order: the central position of the \hi\ source, along with measured heliocentric radial velocity, width, and total flux of the \hi\ detection; the estimated distance to the source from the ALFALFA catalog \citep{haynes18a}; the log of the total mass, column density, and radius of the radio source; the ratio of \hi\ mass to stellar mass; the ratio of \hi\ mass to optical luminosity in the $g-$, $r-$, and $B$-band filters; the peak surface brightness of the optical counterpart in the $g$- and $r$-band; the total apparent and absolute magnitudes of the optical counterpart in the $g$ and $r$ filters; the $g-r$ color of the optical counterpart; the $V$ apparent magnitude and $B-V$ color of the optical counterpart; and the estimated stellar mass and half-light radius (in both arc~seconds and kpc) of the optical counterpart. The optical surface brightness values, magnitudes, and colors have been corrected for Galactic extinction by applying the \citet{schlafly11a} coefficients to the reddening values from \citet{schlegel98a}. The peak \hi\ column density is measured in the highest resolution (5.1\arcsec\ beam) moment 0 map.}
\label{table:properties}
\end{deluxetable*}

\begin{figure*}[t!]
\centering
\includegraphics[width=0.54\textwidth]{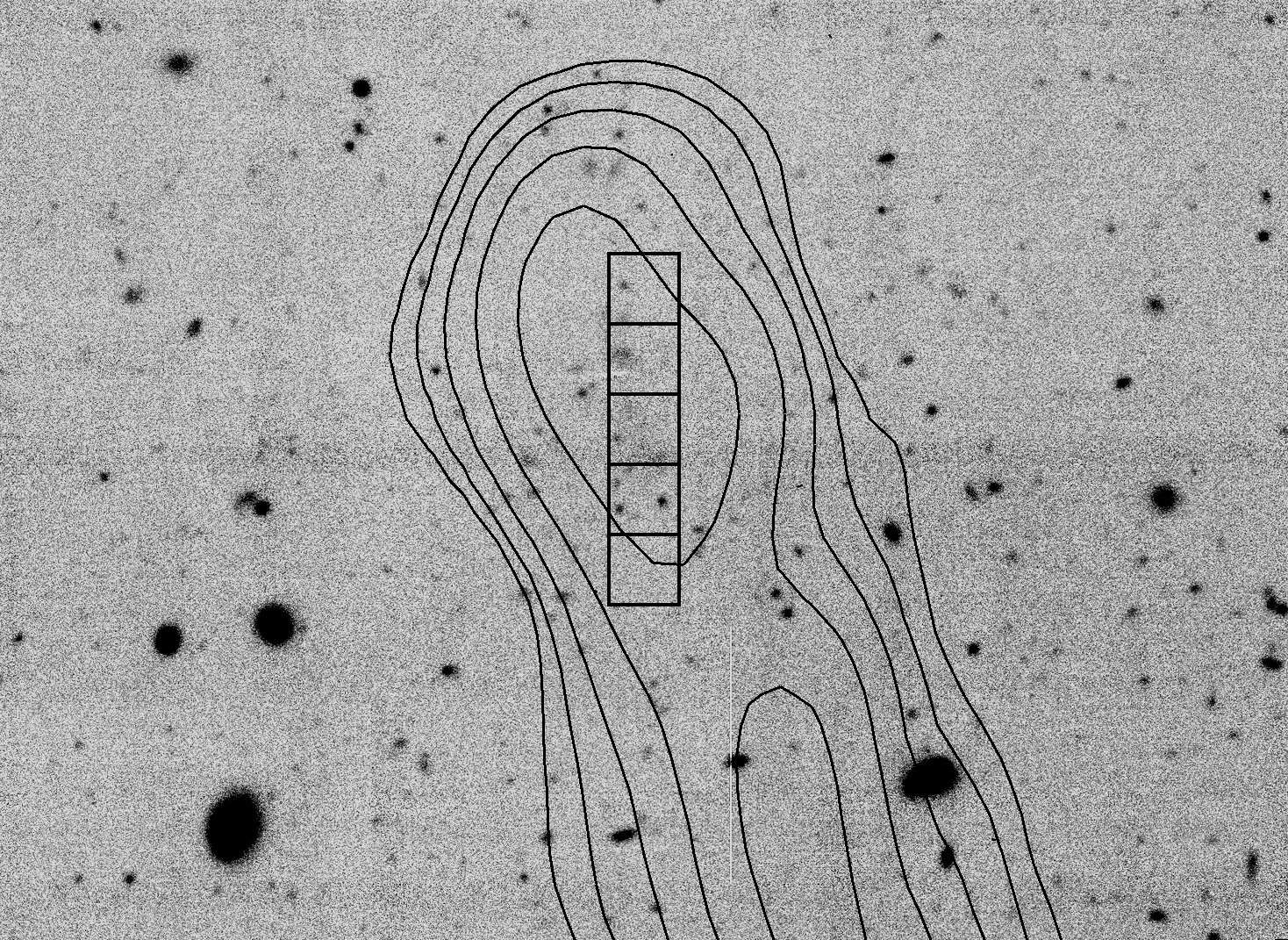}
\includegraphics[width=0.45\textwidth]{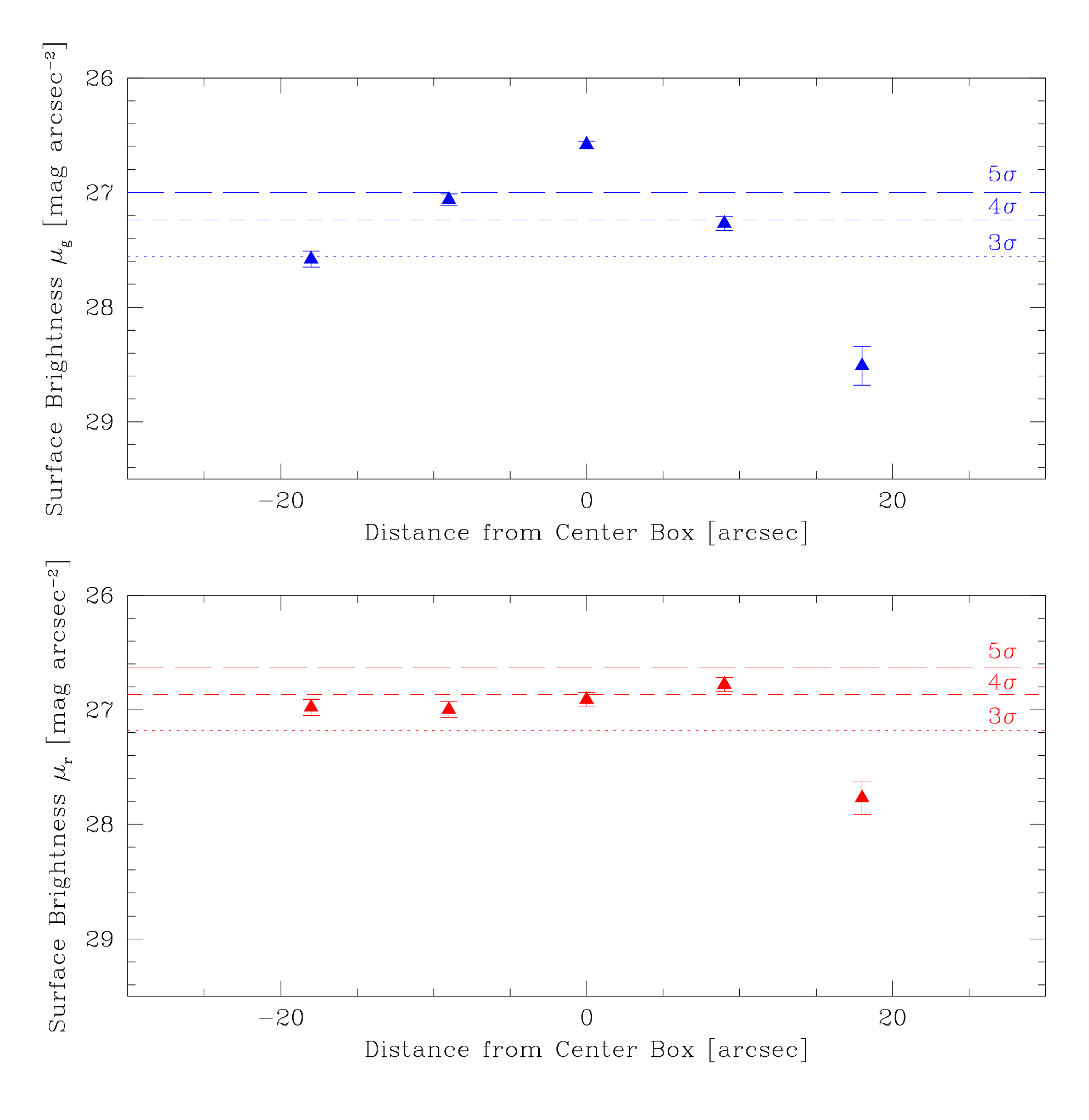}
\vspace{-0.1cm}
\caption{Surface brightness measurements of AGC~229101. The left panel shows a portion of the pODI image that includes the faint optical counterpart that coincides with the northern \hi\ peak; \hi\ contours are again overlaid as in Figure~\ref{fig:main}. 
The five square boxes
used to calculate surface brightness values of the optical counterpart (see Section~\ref{sec:results.optical}) are also shown. 
Two discrete sources that appear to be foreground objects (visible in the fourth box from the top) were masked out before the surface brightness measurements were made. The right panel shows the surface brightness values across the object in the $g$ filter (top) and the $r$ filter (bottom).  
The dotted and dashed lines show the 3-$\sigma$, 4-$\sigma$, and 5-$\sigma$ surface brightness detection thresholds in each filter.
\label{fig:optical}
}
\end{figure*}

\subsection{The Stellar Contents of AGC~229101}
\label{sec:results.optical}

Our deep WIYN pODI imaging reveals an extremely low surface brightness optical counterpart centered at the northern peak of the \hi\ emission (Figure \ref{fig:main}).  This optical source is barely detectable in the WIYN pODI data, as detailed below and illustrated in Figure \ref{fig:optical}. Table~\ref{table:properties} lists the various quantities that we derived from the combined analysis of the WIYN imaging and the \hi\ data; below we describe how the optical quantities were measured and/or calculated. 

\subsubsection{Optical Surface Brightness}
\label{sec:results.optical.sb}

The low S/N nature of the optical source precludes standard surface brightness profile fitting; therefore to measure the surface brightness of AGC~229101, we followed a process similar to that described in \cite{janowiecki15a}, which presented WIYN pODI imaging of the low surface brightness optical counterpart to AGC~228385 (Coma~P). We first determined the local mean sky background level in the images using 20 small boxes, each 45 pixels ($\sim$5\arcsec) on a side, positioned on blank patches of sky around the northern HI clump.  We then placed five square boxes, each 80 pixels ($\sim$9\arcsec) on a side, across the optical emission that appears in the northern HI region (see the left-hand panel of Figure \ref{fig:optical}).  The choice of 80 pixels for the box width was a compromise -- the boxes were large enough that a few (five) boxes could cover the entire area of the optical emission, but small enough to limit the amount of sky background included in each box.  We masked out two sources that were located within one of the object boxes and appeared to be foreground objects separate from the diffuse emission from the optical counterpart. The two sources are quite faint (with $g$ $\sim$ 24.2 and 25.0 mag) and significantly bluer than the emission from the optical counterpart.  Note that Figures~\ref{fig:main} and \ref{fig:optical} show the unmasked images, and the two discrete sources are visible in the fourth box from the top in Figure~\ref{fig:optical}.
We measured the counts per pixel within each of the five boxes and subtracted the local mean sky background level (in counts per pixel).
We then converted the net counts per pixel in each box to a surface brightness level in units of mag~arcsec$^{-2}$.  The measured standard deviation of the sky level of the 20 background boxes was used to calculate the surface brightness level detection thresholds of the images. The 3-, 4-, and 5-$\sigma$ detection thresholds are, respectively, 27.56, 27.24, and 27.00 mag~arcsec$^{-2}$ in the $g$-band image and 27.18, 26.87, and 26.63 mag~arcsec$^{-2}$ in the $r$-band image.   

The results of the surface brightness measurements are plotted in the right-hand panel of Figure \ref{fig:optical}. 
The errors on the surface brightness values take into account the Poisson errors in the counts from the object and the sky, as well as the uncertainty in the measurement of the sky background; the errors range from $\sim$0.03$-$0.06 mag in the central box (where the flux from the optical counterpart is most apparent) to $\sim$0.15-0.17 mag in the outlying boxes. 

The surface brightness of the optical counterpart has a higher S/N level in the $g$-band image -- greater than or equal to 4$\sigma$ in the central three consecutive boxes, 
with the central box reaching more than 0.4 mag~arcsec$^{-2}$ brighter than the 5-$\sigma$ threshold.  The S/N level 
is lower in the $r$-band, but the surface brightness values in the central boxes still hover around the 3- and 4-$\sigma$ thresholds.  
The brightest surface brightness values in $g$ and $r$ are 26.58$\pm$0.03 mag~arcsec$^{-2}$ (in the central box) and 26.78$\pm$0.06 mag~arcsec$^{-2}$ (in the box 9$\arcsec$ north of the central box), respectively. 

Since each box is 9$\arcsec$ square, we estimate the length of the optical counterpart in our images to be approximately 27\arcsec, which translates to $\sim$13~kpc at the 105.9~Mpc distance of AGC~229101.  This is roughly a factor of three smaller than the length (along the major axis) of the northern portion of the \hi\ distribution (see Section~\ref{sec:results.hi}) and about six times smaller than the total (north-to-south) extent of the \hi. 

\subsubsection{Total Magnitude, Color, and Radius}
\label{sec:results.optical.magrad}

We calculated the global properties of the optical counterpart by carrying out aperture photometry of the object on the $g$ and $r$ images.  To ensure that we were measuring an accurate total magnitude,
we measured the flux within a large aperture that was positioned and sized so that it included the three consecutive central boxes, all of which showed a significant detection in the surface brightness measurements.  The center of the aperture also coincides with the \hi\ centroid position of the northern \hi\ column density peak. The aperture has a diameter of 27\arcsec\ (246 pixels). The optical counterpart is so faint that the flux from the sky background dominates the measurement; thus we also measured the flux at the same position but using a much smaller aperture, with a diameter of 12\arcsec\ (106 pixels), in order to try to limit the contribution from the sky background and perhaps measure a more representative color for the object. 

The total magnitudes and colors 
are listed in Table \ref{table:properties}. The $g$ and $r$ magnitudes in the table are those derived from the 
larger aperture, since that aperture encompasses the emission from the optical counterpart that is apparent in the image.  We list the $g-r$ color for both the larger and smaller aperture, in order to document the effect of the aperture size on the measured color. 
While the optical counterpart has a blue color in both 
apertures, it ranges from $g-r$ $=$ 0.44$\pm$0.26 mag within the larger aperture to as blue as 0.06$\pm$0.13 mag in the smaller aperture.  We use the relations given in \citet{jester05a} to convert our $g$, $r$ photometry results to $V$ and $B-V$ values and list them in Table~\ref{table:properties}. 
Combining the total $V$ magnitude with the distance modulus for AGC~229101 yields an absolute $V$-band magnitude of $M_V$$=$ $-$13.38 mag.

Next we derived an empirical half-light radius for the optical counterpart.  First, we determined the total number of sky-subtracted counts contained within the large 
aperture that was used to calculate the total apparent magnitudes.
We then carried out photometry of the object with a series of apertures, starting with a small radius and increasing the aperture radius by one pixel (0.11\arcsec) each time.  We used the results of these measurements to calculate the aperture radius that encloses half of the total counts, and adopt that as the  half-light radius. 
The estimated half-light radius, in both arc~seconds and kpc, is listed as r$_{h}$ in Table~\ref{table:properties}. The uncertainty in this quantity takes into account the uncertainty in the location of the half-light radius and the uncertainty on the total flux from the object. 

\subsubsection{Stellar Mass}
\label{sec:results.optical.smass}

Stellar mass estimates are challenging in that they rely on a detailed understanding of the underlying stellar population in a given galaxy. 
In the absence of spectroscopic data, most photometric estimates rely on calibrated color -- stellar mass-to-light ratio relations (CMLRs) to estimate stellar mass-to-light ratios. While CMLRs have been calibrated on model data, and a range of observational samples (see, e.g., \citealp{du20b} and references therein), 
the applicability of these fits to galaxies like AGC~229101 with extreme or unusual properties is not clear. Still, we apply a selection of CMLRs in order to place a constraint on the range of stellar masses and to facilitate comparison to other work.

To do this we combine 
the total $g-r$ color from the large-aperture photometry with three different CMLRs 
to estimate stellar mass-to-light ratios in the $g$- and $r$-bands, and then combine these with the large-aperture stellar luminosity in the corresponding filter to compute total stellar mass. Though there are many available measurements and calibrations of the CMLR, we choose those from \citet{bell03a}, \citet{herrmann16a}, and \citet{du20a} -- which are calibrated with spiral, dwarf, and low surface brightness galaxies respectively -- to attempt to encompass the range of potential underlying stellar populations relevant to AGC~229101. The \citet{du20a} relations give the lowest estimated stellar masses (1.31 \& 1.35$\times10^7\msun$), and the \cite{bell03a} relations give the highest values (2.87 \& 2.86$\times10^7\msun$). The \citet{herrmann16a} relation, which was applied to a sample of ALFALFA UDGs in \cite{mancerapina19a}, gives 1.80$\times10^7\msun$. All three relations also give slightly different values depending on the chosen optical band. 

The large uncertainty in our color measurement creates additional uncertainty in our stellar mass estimate, further complicating the picture.
Simply varying the color by adding or subtracting the associated error produces a substantial change in the calculated mass-to-light ratio, which in turn yields a total stellar mass that varies significantly, by a factor of two or more, for all three relations. 

Since it is not clear which relation ultimately best applies to the stellar population of AGC~229101, we take the mean of the results
, 2.1 $\times$ 10$^7$ solar masses ($10^{7.32}~\msun$), as the final stellar mass for the optical counterpart.

Taking into account the color uncertainty, as well as the uncertainty in the total magnitudes and the spread in values associated with utilizing different CMLRs, we estimate the uncertainty on the stellar mass to be $\pm$0.33~dex (see Table \ref{table:properties}).   

As an alternative approach, we note that perhaps the most in-depth study of the stellar populations of gas rich ALFALFA galaxies to date is \cite{huang12b}. They estimate stellar masses for all ALFALFA galaxies that fall in the GALEX footprint using UV-optical SED fitting following \cite{salim07a} and \cite{brinchmann04a}. These stellar mass estimates are systematically lower than, e.g., \cite{bell03a} by 0.53~dex, but represent a better calibration of gas-rich, nearby star-forming galaxies, with improved estimates of their internal extinction. While we do not have a full SED for AGC~229101, we fit the offset between the CMLR masses and SED masses for the ALFALFA sample, and then apply the standard offset between the two samples to AGC~229101, giving a log stellar mass of 6.93 solar masses, 
slightly lower than our estimate from the CMLRs above. We use this value when directly comparing to the \cite{huang12b} sample in Section~\ref{sec:discussion.extreme}.

\subsubsection{Ratio of HI Mass to Stellar Luminosity and Mass}
\label{sec:results.optical.ratios}

We calculated \hi\ mass to stellar luminosity and mass ratios using the total \hi\ mass, which includes both the northern and southern components, and report the values in Table~\ref{table:properties}. The \hi\ mass to $g$ luminosity, M$_{\rm HI}$/L$_g$, and $r$ luminosity, M$_{\rm HI}$/L$_r$ are calculated converting the measured absolute magnitudes to luminosity using AB solar magnitudes from \cite{willmer18a}.
To calculate a value for the ratio of the \hi\ mass to the $B$-band luminosity for AGC~229101, we combine the total $V$ magnitude of the object (calculated as described above, using the measured $g$ and $g-r$ values and the conversion equation from Jester et al.\ (2005)) with the $B-V$ color measured from the larger aperture to calculate the total apparent $B$ magnitude.  At the 105.9~Mpc distance of AGC~229101, this yields an absolute $B$-band magnitude of $-$12.73 and combining this with the M$_{\rm HI}$ value gives a ratio of the \hi\ mass to $B$-band luminosity of 110. The errors on the \hi\ mass to optical luminosity ratios that are listed in Table~\ref{table:properties} reflect the range of values we get for these quantities if we vary the optical luminosity values according to their corresponding photometric measurement errors, since these photometric errors dominate the overall uncertainties. 
Combining the measured \hi\ and optical masses yields an \hi\ to stellar mass ratio of 98 
for AGC~229101.  Note that to calculate this ratio, we are using the total \hi\ mass, which includes both the northern and southern components.  Because the stellar mass is so uncertain (10$^{7.32 \pm 0.33}$ M$_\odot$)
, this ratio could vary from as small as 46 
to as large as 209
. 
We noted in Section~\ref{sec:results.hi} that the northern and southern portions of the \hi\ distribution are approximately equal in mass, and 
that the location of the optical counterpart coincides with the northern \hi\ lobe.  Therefore it may be appropriate to calculate quantities like the \hi-to-stellar mass using only the northern portion of the \hi\ gas, and assume that the southern portion of the \hi\ source is an optically dark "tail" of neutral hydrogen.   If one considers only the \hi\ gas in the northern lobe, then all of the quantities that combine the \hi\ and optical measurements (namely the \hi\ to stellar mass ratio and the \hi\ mass to stellar luminosities listed in Table~\ref{table:properties}) would be reduced by a factor of two (to M$_{\rm HI}$/L$_g\approx$ M$_{\rm HI}$/L$_r\approx$ 53~$\msun/\lsun$ and M$_{\rm HI}$/M$_{\star}$ = 49).  

\subsubsection{UV Data}
\label{sec:results.optical.uv}

We searched the GALEX\footnote{Based on observations made with the NASA Galaxy Evolution Explorer. GALEX is operated for NASA by the California Institute of Technology under NASA contract NAS5-98034.} archive for observations of AGC~229101 and found that the area was included in the All-sky Imaging Survey (AIS).  We examined shallow (107-second exposure time) NUV and FUV imaging that included  AGC~229101's position, with the location of the object roughly 3~arc minutes from the edge of the GALEX pointing. No detectable UV emission appeared in the image at that location.

\subsection{AGC~229101: Environmental Context}
\label{sec:discussion:environment}
AGC~229101 is located in a filamentary region of the Coma Supercluster, midway between the Coma Cluster (Abell~1656) and the Leo Cluster (Abell~1367). It lies just over 2 degrees ($\sim$3.5 Mpc) in projection to the northeast of two large groups, the NGC~4213 group and the CGCG~128-034 group.  
Evaluation of the locations of galaxies with available redshifts on larger scales indicates that AGC~229101 is situated in a region of modest overdensity. 

\begin{figure*}[t!]
\begin{minipage}{.66\textwidth}
  \centering
  \hspace{-0.1cm}
\includegraphics[width=1.05\textwidth]{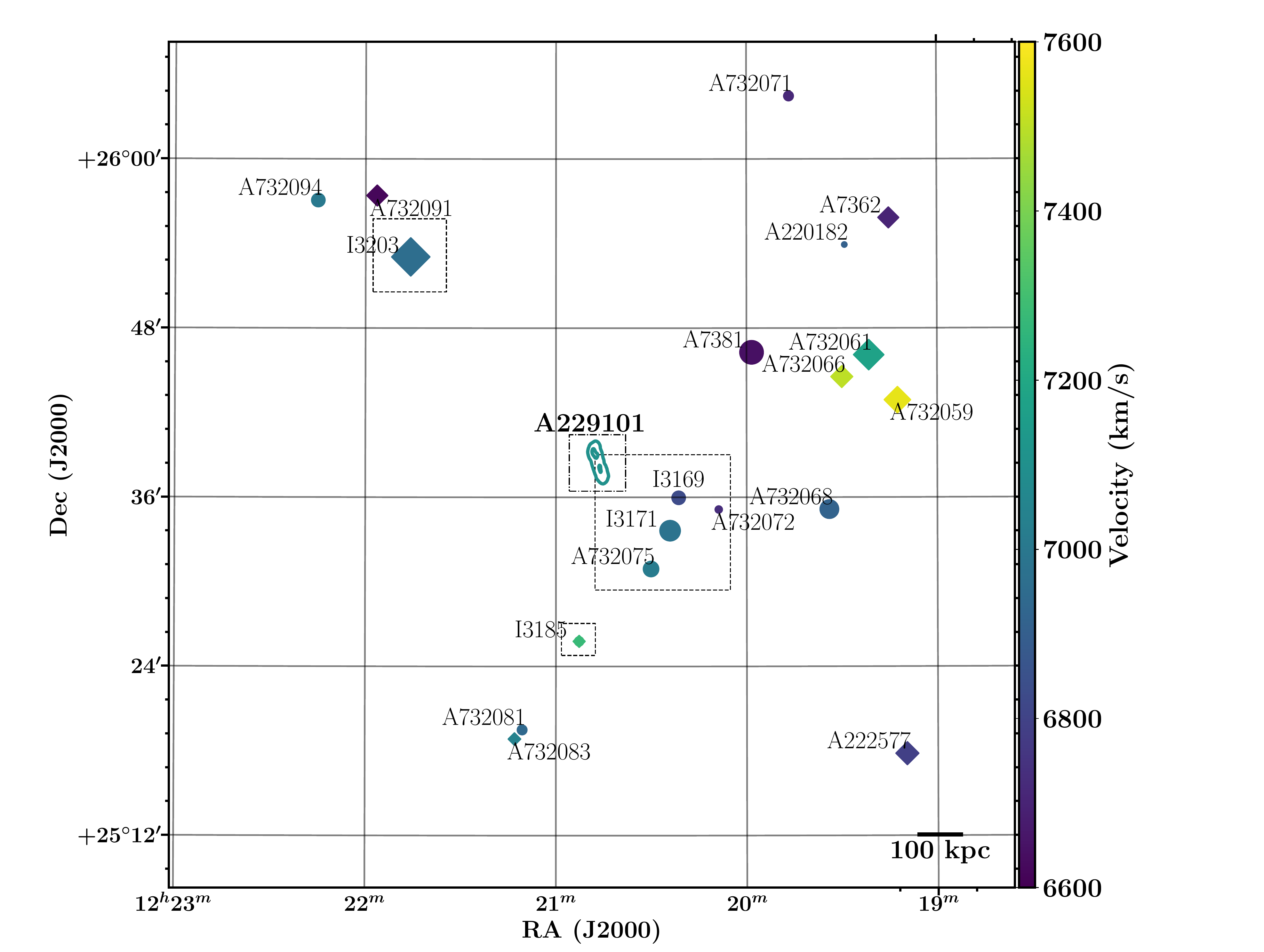}
\end{minipage}
\begin{minipage}{.34\textwidth}
  \centering
\includegraphics[width=1.0\textwidth]{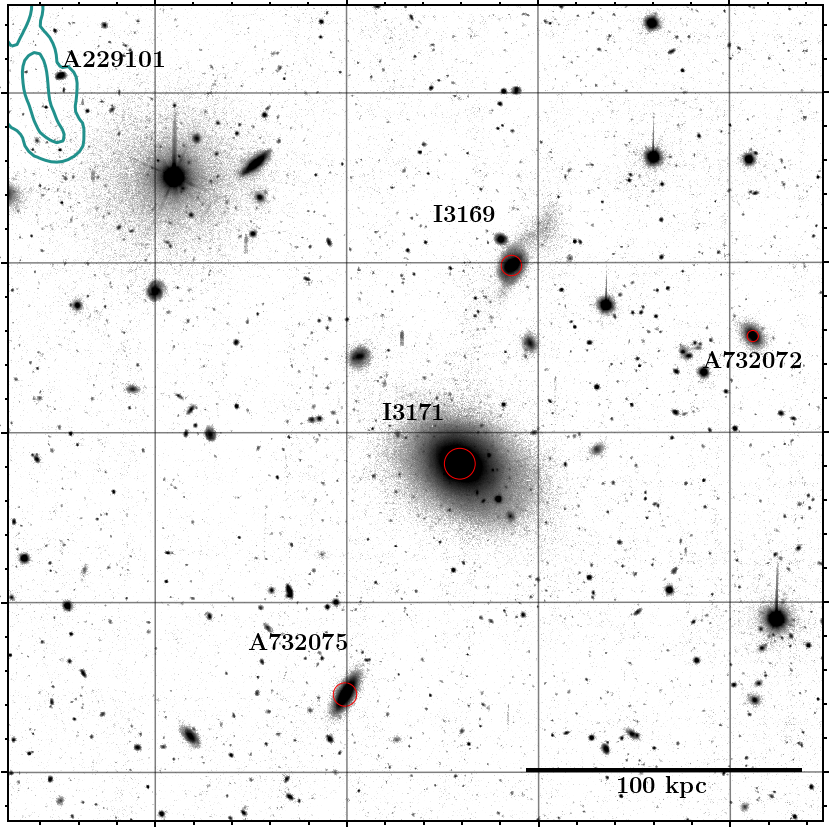} 
 \includegraphics[width=0.5\textwidth]{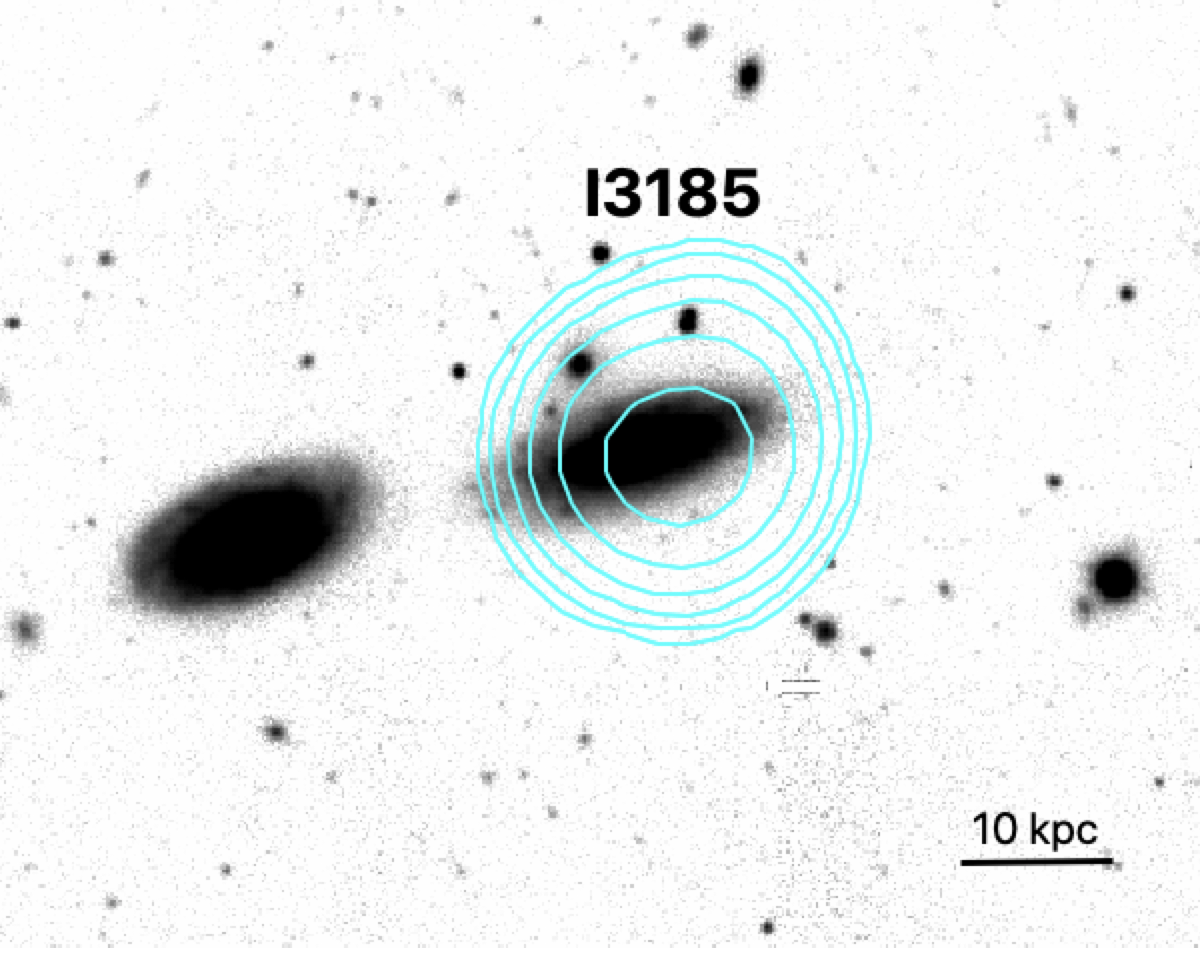} 
  \includegraphics[width=0.48\textwidth]{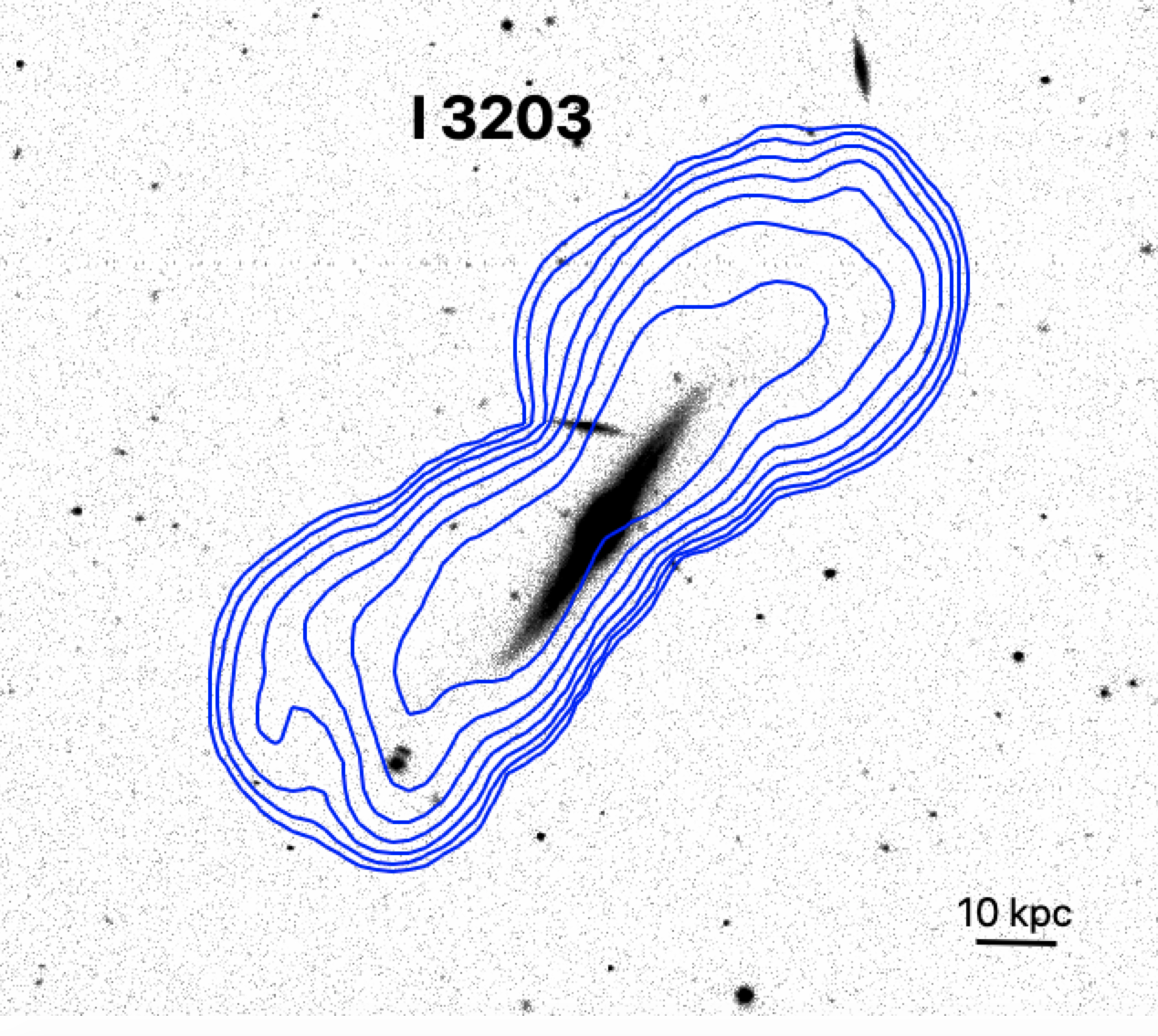} 

\end{minipage}
\vspace{-0.1cm}
\caption{Left: Shown here are galaxies within a projected angular separation of $\sim$0.5-degree ($\sim$1~Mpc at a distance of 105.9~Mpc), and a recession velocity within 500~\kms, of AGC~229101's position and velocity. Some of these objects may be possible tidal companions to AGC~229101. The size of each symbol scales with the $z$-band magnitude of each source, and colors indicate measured recessional velocities as shown in the color bar. Sources with detected \hi\ are marked with diamonds and sources without an \hi\ detection are marked with circles.  AGC~229101 is plotted with its \hi\ contours at the center of the figure. The region depicted in Figure \ref{fig:main} is indicated with a dash-dot line, and the regions depicted in the right hand panels are depicted with dashed lines.
Right Top: pODI image of IC~3171 and surrounding galaxies to the southwest of AGC~229101. The location of AGC~229101 is indicated by its \hi\ contours in the upper left corner of the figure.
Right Bottom: pODI image of IC~3185 with WSRT contours overlaid in light blue, and SDSS image of IC~3203 with WSRT contours overlaid in blue. Note that IC~3203 is outside the FWHM of the WSRT primary beam, but is still bright enough to be detected. WSRT contours for both images range from 1-32$\times10^{19}$atoms~cm$^{-2}$ in powers of 2, and have been primary beam corrected.
\label{fig:environment}
}
\end{figure*}

Figure \ref{fig:environment} shows all galaxies that have measured redshifts within 500 \kms\ of AGC~229101's redshift and are located within a projected angular separation of $\sim$0.5~degree (which corresponds to $\sim$1~Mpc at 105.9~Mpc) from the source. AGC~229101 appears to be on the northeastern edge of a loose collection of galaxies that are located at similar distances ($\sim$100~Mpc). It is projected on the sky $\sim$200~kpc to the northeast of its nearest neighboring galaxies, IC~3169 and IC~3171 (CGCG~128078), with a recessional velocity that is 286 \kms\ and 155 \kms\ larger than the velocity of those objects, respectively.  
Both IC~3169 and IC~3171 appear to have early-type morphologies, and \citet{consolandi16a} estimate their stellar masses to be log~$(M_{\star}/M_{\odot})=9.93$ and log~$(M_{\star}/M_{\odot})=10.85$, respectively.  
Neither galaxy is detected in \hi\ at the sensitivity of ALFALFA, but both are visible in the WIYN pODI image. 
IC~3169 exhibits extended low surface brightness emission to the northwest that has the appearance of a tidal feature,
while IC~3171 has an unperturbed elliptical light distribution with no obvious evidence of disruption.
 
 The closest galaxy to AGC~229101 with detected \hi\ is IC~3185 (AGC~225885), which is located to the south at a projected separation of 390 kpc 
 and which has a recession velocity within 100 \kms\ of that of AGC~229101. IC~3185 is a gas-rich star-forming galaxy with a stellar mass of $M_{\star}/M_{\odot}=9.39$ and an \hi\ mass of 
 $M_{\rm HI}/M_{\odot}=9.52$ \citep{consolandi16a,haynes18a}. It has a nearby (in projection) companion (IC~3189) that is actually a background source with a recession velocity of 15283 \kms 
 \citep{sanchezalmedia11}.
 
 Farther away and significantly more massive than IC~3185 is the galaxy IC~3203.  This edge-on spiral galaxy is located 610~kpc in projection to the northeast of AGC~229101. With a stellar mass of $M_{\star}/M_{\odot}=10.45$ 
 and an \hi\ mass of $M_{\rm{HI}}/M_{\odot}=10.11$ \citep{consolandi16a,haynes18a}, IC~3203 is the most massive \hi-bearing galaxy in the vicinity of AGC~229101.   

\section{Discussion}
\label{sec:discussion}

To summarize the results in the previous sections, AGC~229101 has a massive and extended \hi\ distribution and a much less extended optical counterpart that coincides with the northern portion of the \hi\ gas component. Its $\sim$2 x 10$^9$ solar masses of \hi\ (compared to roughly $\sim$5 x 10$^9$ for the Milky Way; \citealp{henderson82a}), 
extends $\sim$80~kpc north-to-south across the sky in two main portions, with a peak column density around $\sim$5 x 10$^{20}$ atoms~cm$^{-2}$ at our highest resolution ($\sim$2.8~kpc beam). The \hi\ line width at the 50\% flux level across the full source is $<$50~\kms, with widths of 31 \kms\ and 41 \kms\ for the northern and southern clumps respectively. A possible optical counterpart 
is barely detectable in our deep optical imaging, with a peak surface brightness fainter than 26.5 mag/arcsec$^{-2}$ in both the $g$ and $r$ bands.  The optical counterpart coincides with the northern \hi\ peak and has a blue color and an absolute magnitude 
consistent with the color and luminosity of a dwarf galaxy.  
We estimate that the stellar mass of the optical counterpart is 10$^{7.32 \pm 0.33}$ M$_\odot$
; the large uncertainties on the stellar mass are due to the faintness of the optical emission.  

The observed properties of AGC~229101 -- specifically, the object's high gas fraction, low optical surface brightness, and low velocity width given its \hi\ mass -- make it stand out compared to other sources in the ALFALFA survey.   The reason for these extreme properties is not immediately clear. Here we consider 
AGC~229101 in the context of other sources in the ALFALFA survey and the literature, and then explore a number of hypotheses for the exceptional nature of this object. 

\subsection{The Extreme Nature of AGC~229101}
\label{sec:discussion.extreme}


\begin{figure*}[t!]
\centering
\includegraphics[width=0.325\textwidth]{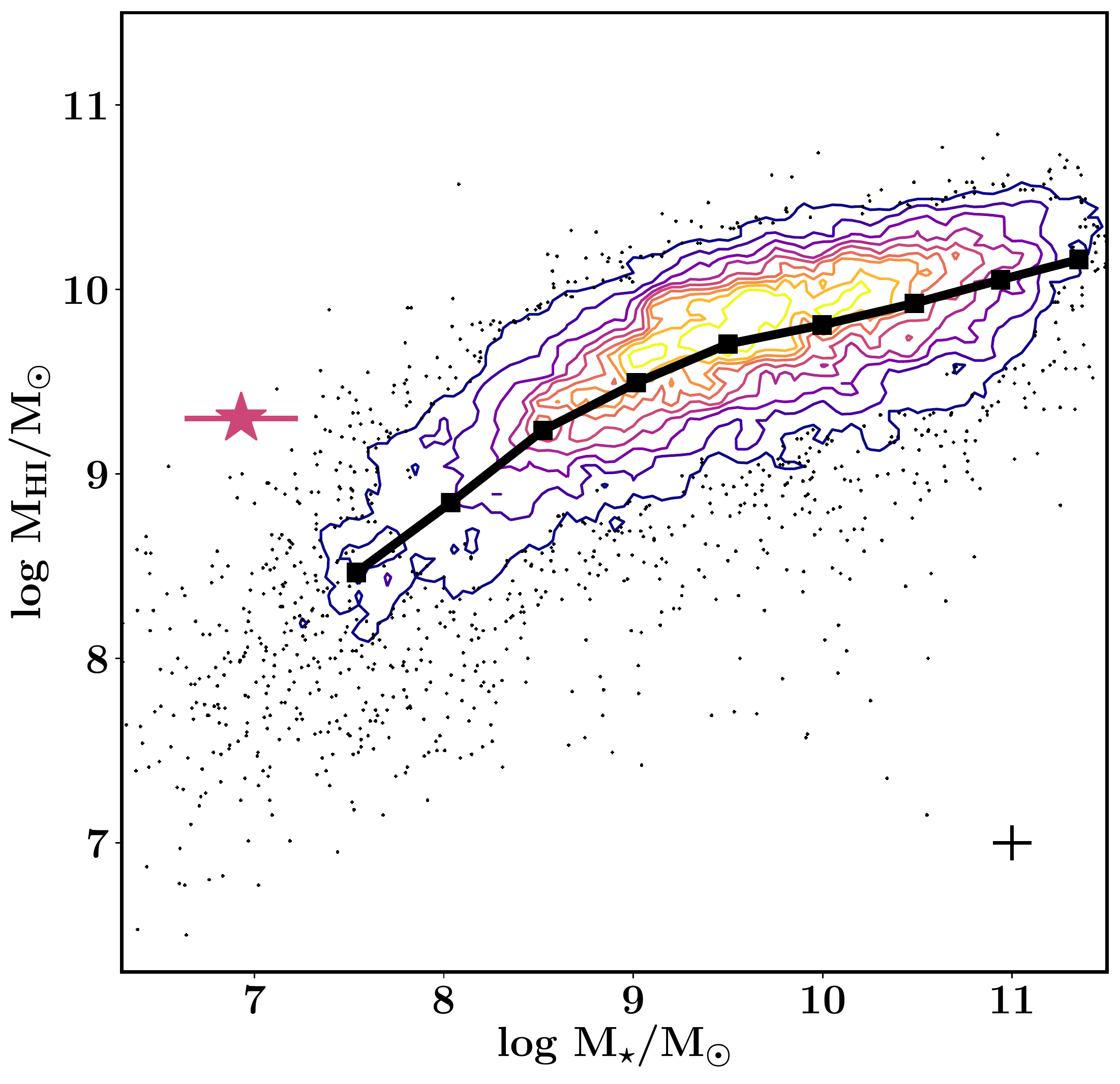}
\includegraphics[width=0.335\textwidth]{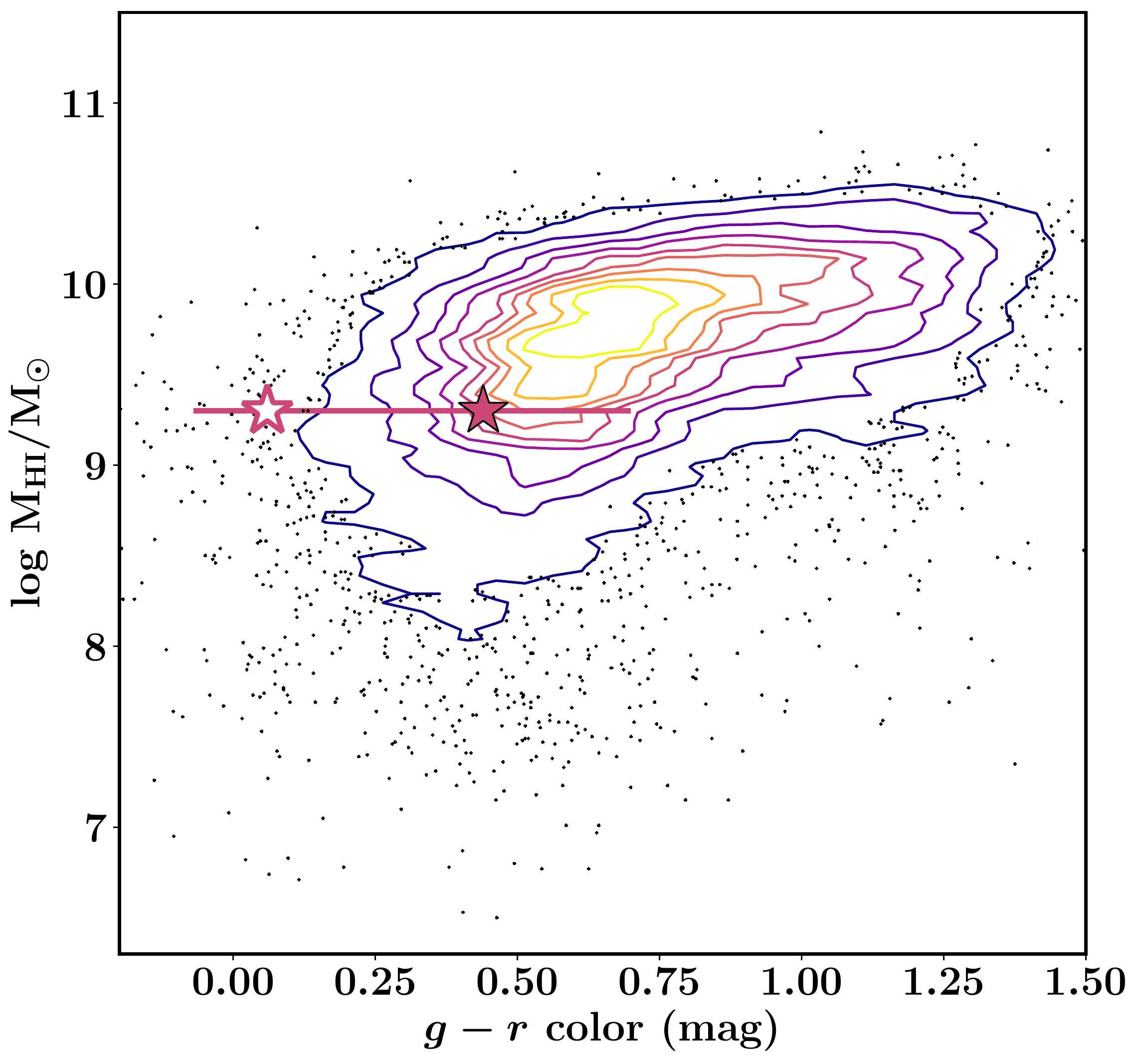}
\includegraphics[width=0.328\textwidth]{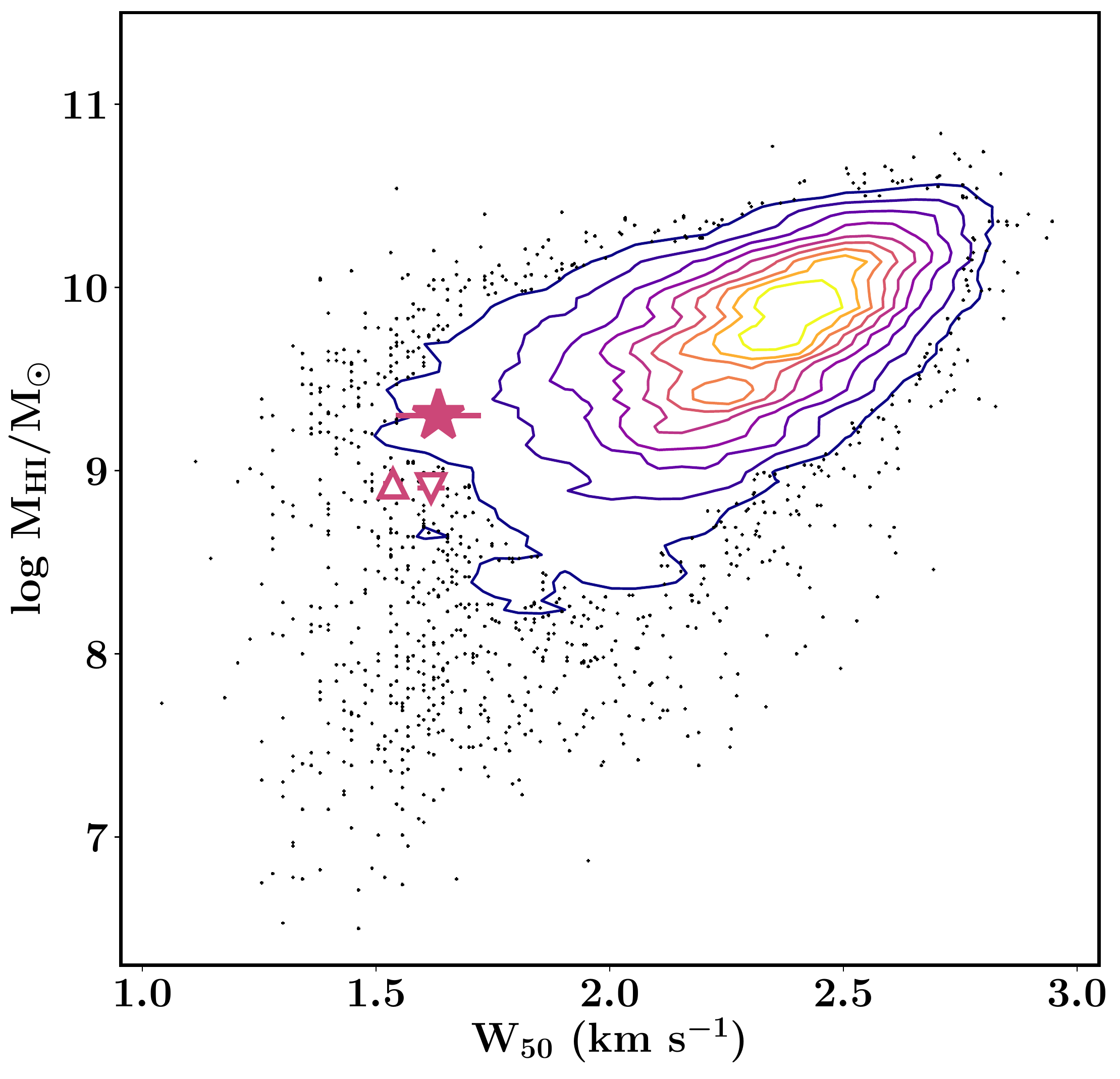}
\caption{AGC 229101 compared with ALFALFA galaxies from \cite{huang12b}. Contours represent ALFALFA sources as measured by \cite{huang12b} in 10\% intervals from 10\% to 90\%, with outliers shown by grey dots. 
Left: \hi-mass -- stellar mass relation for ALFALFA sources, with stellar masses derived from SED fitting best suited to nearby, gas rich galaxies, as described in \cite{huang12b}. The median relation is shown in black, and typical uncertainties are shown by a black cross in the lower right corner. AGC~229101 is shown as a filled star symbol far off the relation. Note that the stellar mass for AGC~229101 plotted here is the stellar mass derived by matching to the \cite{huang12b} masses as discussed in Section \ref{sec:results.optical.smass}. Further note that only including gas in the northern component of AGC~229101 reduces the \hi\ mass by 0.3~dex, but still is extreme compared to extrapolation of the median ALFALFA sample. Center: \hi-mass versus g--r color for ALFALFA galaxies, compared with AGC~229101.  The two measured colors (for two different apertures; see Section~\ref{sec:results.optical.magrad}) are shown as filled and unfilled symbols and plotted with error bars, which overlap.
Right: \hi-mass versus \hi\ line width measured at the 50\% flux level. The upward pointing triangle represents the northern \hi\ peak, and the downward pointing triangle the southern \hi\ peak. The line width for A229101 -- the entire source and the individual peaks -- is very narrow relative to the ALFALFA sample.
\label{fig:comparison}
}
\end{figure*}

The ALFALFA survey has detected a number of sources that have some of the same characteristics that AGC~229101 exhibits -- e.g., sources that have very low surface brightnesses and extended half-light radii for their stellar masses (dubbed HUDs, or HI-bearing Ultra-Diffuse sources; see \citealp{leisman17a}), and sources with highly elevated gas fractions (see, e.g., \citealp{cannon15a, janowiecki15a}).  AGC~229101 is in some ways more extreme than even these objects.  
Figure \ref{fig:comparison} shows the global properties of AGC~229101 compared with the corresponding properties of the ALFALFA \hi-selected population, as measured by \citet{huang12b} (we choose to use this comparison sample since it represents the most carefully measured estimates of ALFALFA stellar masses to date - see Section \ref{sec:results.optical.smass}).  
The left panel shows a plot of \hi\ mass versus estimated stellar mass.  AGC~229101 is marked with a star that lies far from the median relation, demonstrating the extreme gas-rich nature of AGC~229101 compared with other ALFALFA detections.   The average \hi\ mass of an ALFALFA-detected source with a stellar mass near $\sim10^7\msun$ (i.e., the stellar mass of AGC~229101 when scaled to match the \cite{huang12b} estimates) is expected to be $\sim 10^8\msun$, approximately an order of magnitude lower than the observed \hi\ mass of AGC~229101. This result holds even if we consider only the northern component, which has an \hi\ mass of $10^{9.01}\msun$, and is especially significant given that the comparison sample is measured using SDSS catalog photometry, which can over-represent the scatter in the most extreme points due to photometric issues like poor background subtraction near bright stars.
We note also that the comparison sample, which is from the \hi-selected ALFALFA survey, will tend to be gas-rich relative to typical galaxies (see, e.g.,  \citealp{catinella10a,bradford15a}). AGC~229101 has an \hi\ mass to $B$-band stellar luminosity ratio $M_{HI}/L_B$ of 110, compared with typical values which range between $\sim$0.1 and 4 for dwarf and irregular galaxies. \citep{roberts94a,stil02a,lee03a}. 

The center panel of Figure \ref{fig:comparison} shows the g--r color of AGC~229101 estimated from our WIYN pODI imaging compared with the g--r colors of ALFALFA galaxies measured by SDSS. AGC~229101 is relatively blue, even when compared to this already blue-biased \hi-selected comparison sample. 

The right hand panel of Figure \ref{fig:comparison} shows the velocity width of the overall source, and the northern and southern components compared with the ALFALFA galaxies, demonstrating its narrow line width compared to the ALFALFA sample for similar HI masses. We note that while this source may not be a disk, its elongated nature, and the elongated nature of both components, suggest that the low line width is not simply an inclination effect. In fact, if its \hi\ axial ratio is representative of its inclination, then AGC~229101 falls off the Baryonic Tully Fisher relation, consistent with having no ``missing baryons'' (with a baryon fraction approximately or greater than the cosmological baryon fraction; see  \citealp{mancerapina19a}).

In addition to its high gas fraction, blue color, and narrow line width, the low surface brightness nature of AGC~229101 is striking. AGC~229101 is among the lowest surface brightness extragalactic sources in the sample of $>$30,000 galaxies detected by ALFALFA. 
Indeed, the peak surface brightness of AGC~229101 is well below the typical cutoff that defines the isophotal edge of a galaxy, 25 mag~arcsec$^{-2}$, and is $\sim$1.6 mag~arcsec$^{-2}$ fainter than the median central surface brightness of the already extreme UDGs \citep{vandokkum15a} (see Section~\ref{sec:discussion:UDG} for more discussion). 

\begin{figure*}[t!]
\centering
\includegraphics[width=0.9\textwidth]{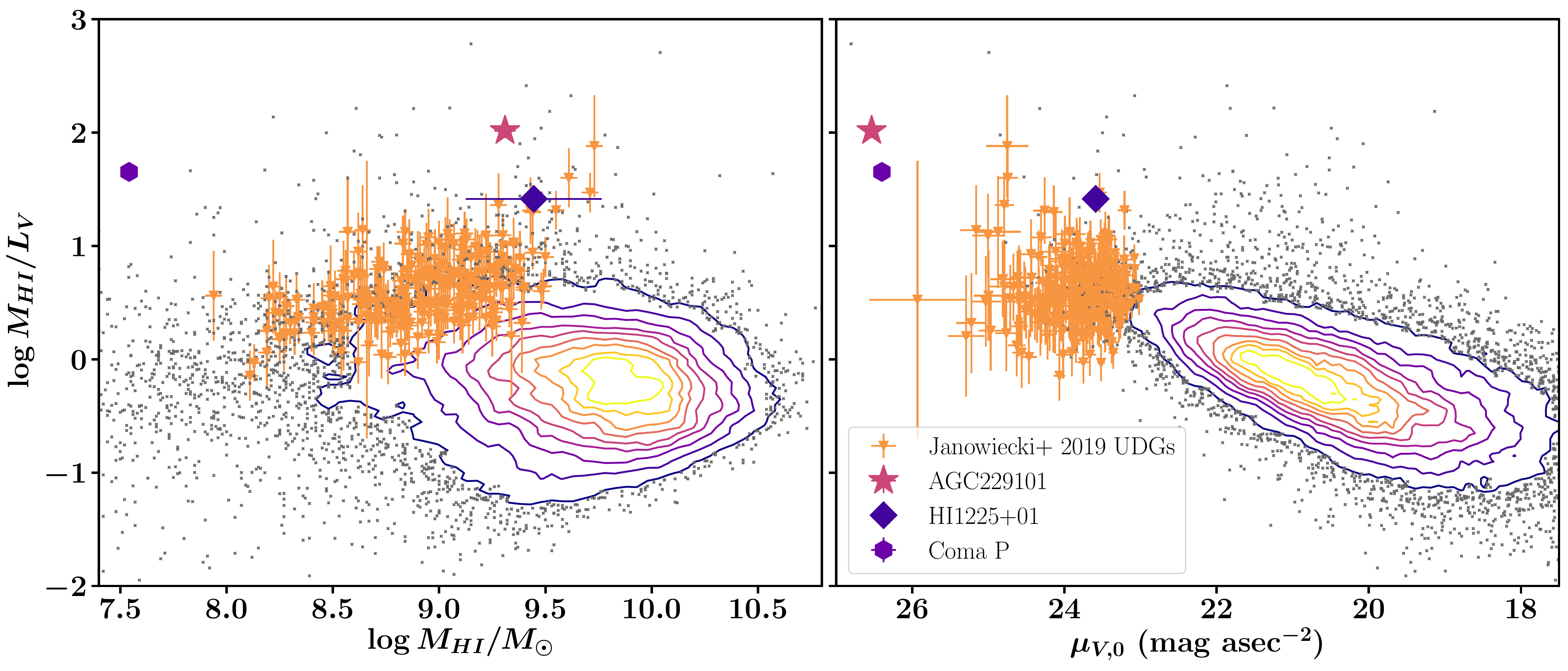}
\caption{AGC 229101 compared with ALFALFA \hi-bearing ultra-diffuse galaxies, and other extreme sources, demonstrating the anomalous  nature of AGC~229101 in this parameter space. The panels show the \hi-mass -- V-band luminosity ratio vs \hi-mass (left) and V-band peak surface brightness (right).  Contours represent ALFALFA sources as measured with SDSS photometry (see, e.g., \citealp{haynes11a,haynes18a}) increasing in 10\% intervals, with outliers shown with small grey points (note: the most extreme outliers have poor optical photometry, as discussed in Section \ref{sec:discussion.extreme}). 
ALFALFA Ultra-diffuse galaxies from \citealp{leisman17a} and \citet{janowiecki19a} are shown with orange triangles, and the extreme ALFALFA sources Coma P and HI1225+01 discussed in the text are shown as a purple hexagon and blue diamond respectively. AGC~229101 is shown as a filled pink star, far lower surface brightness and higher gas fraction than almost all other ALFALFA galaxies. Note that the error bars on AGC~229101 are the size of or smaller than the marker.
\label{fig:comparison2}
}
\end{figure*}

As shown in the right hand panel of Figure \ref{fig:comparison2}, the faint peak surface brightness, high gas fraction, and fairly blue color of AGC~229101 are reminiscent of those of the aforementioned Almost-Dark galaxy Coma~P (AGC~229385). Coma~P has a peak surface brightness of $\mu_{g}$ $=$ 26.4~mag~arcsec$^{-2}$, very close to that of AGC~229101, and a $g-r$ color of $-$0.05$\pm$0.09 \citep{janowiecki15a,brunker19a}, which is bluer than the color we measure in the large aperture for AGC~229101 ($g-r$ $=$ 0.44$\pm$0.26 mag), but comparable to the color we measure in the smaller aperture ($g-r$ $=$ 0.06$\pm$0.13 mag).  Coma~P is likewise one of the most gas-rich systems in the ALFALFA catalog, with an \hi-to-stellar-mass ratio $M_{\rm HI}/M_{*}$ of 81 \citep{brunker19a}, compared to $M_{\rm HI}/M_{*}$ $=$ 98 
for AGC~229101.  However, this comparison is strongly dependent on the color used to estimate the stellar mass; the more directly observed $M_{\rm HI}/L_{B}$=110 is nearly 4$\times$ larger than Coma P. Moreover, Coma~P is much smaller than AGC~229101.  Coma~P has 
an \hi\ mass of $\sim3.5 \times 10^7 \msun$, and a stellar mass of $\sim4 \times 10^5 \msun$ \citep{brunker19a}, which means its baryonic mass is over 60 times less than AGC~229101. 
Thus, though AGC~229101 shares many similar properties with Coma P, it is ultimately a much larger, very different source.

Figure \ref{fig:comparison2} also shows AGC~229101 compared to the famous multi-component extreme ``dark" gas cloud in the direction of the Virgo Cluster, HI~1225+01 \citep{giovanelli89a, chengalur95a} -- another ALFALFA source that shares multiple similarities with AGC~229101. 
HI~1225+01 was discovered serendipitously during Arecibo Observatory \hi\ line observations by \citet{giovanelli89a} and its \hi\ distribution was mapped with follow-up observations by \citet{giovanelli91a} and \citet{chengalur95a}. 
Its estimated distance is $\sim$16~Mpc, based on its recession velocity and a flow model \citep{giovanelli91a}, though this is highly uncertain given that it is in the direction of the Virgo Cluster, where flow model distances can be unreliable. 
HI~1225+01 has a similar \hi\ mass to AGC~229101, a similarly large gas fraction, and, perhaps most strikingly, a morphologically similar multi-component \hi\ distribution.
The source has two \hi\ components, with a total \hi\ mass of approximately $2\times10^9$ solar masses, similar to the \hi\ mass of AGC~229101. Moreover, the northeast clump of HI~1225+01 has a low surface brightness, blue optical counterpart while the other clump has no identified stellar emission 
\citep{salzer91a,matsuoka12a}.

However, though broadly similar, the specific properties of the two sources show important differences. The optical counterpart to HI~1225+01 is  bluer in color than AGC~229101 (with $g-r$ $\sim$ 0; \citep{matsuoka12a}), and it has active star formation \citep{salzer91a} and is much more compact
and likely less massive than AGC~229101, with a stellar mass of approximately $10^6$ solar masses \citep{matsuoka12a}. 
Furthermore, the \hi\ distribution of HI1225+01 has an even larger physical extent than does the \hi\ gas in AGC~229101; in HI1225+01, the centers of the two clumps are separated by about 100~kpc and stretch over nearly 200~kpc. On the other hand, given its highly uncertain distance, if HI1225+01 is actually at a closer distance (as Coma~P turned out to be), its physical extent could be more comparable to that of AGC~229101.

\subsubsection{Comparing AGC~229101 to Ultra-Diffuse Galaxies}
\label{sec:discussion:UDG}

In a search for low surface brightness features in the Coma Cluster, \citet{vandokkum15a} identified a population of objects they dubbed ``ultra-diffuse galaxies" (UDGs) that have very low peak surface brightnesses (central surface brightnesses of $\mu_g$ $>$ 25 mag~arcsec$^{-2}$) and large effective radii ($r_{e}>$1.5~kpc) for their stellar masses, which are comparable to the stellar masses of dwarf galaxies ($\sim10^7$ to $10^8 M_{\sun}$).  
Objects that seem to fit the general definition of UDGs have been identified both in dense environments (e.g., \citealp{vandokkum15a,koda15a,mancerapina19a})
as well as in lower-density regions (e.g., \citealp{roman17b,leisman17a,greco18a}.
This would seem to suggest that they represent a heterogeneous population of galaxies that are united by their properties but not necessarily by a common formation mechanism.  Moreover, classes of galaxies with the properties of UDGs -- namely, low peak surface brightnesses, along with large sizes for their stellar masses -- have long been known, and different types of galaxies can of course overlap in terms of their properties \citep{conselice18a}; \citet{vandokkum15a} explicitly argued in their paper that the UDGs they had identified did not represent a distinct population of galaxies, but instead were likely ``the largest and most diffuse objects in a continuous distribution". In any case, it seems relevant to compare the observed properties of AGC~229101 with those of other low surface brightness, diffuse, low-mass galaxies. 

With a peak surface brightness $\mu_g$ of 26.6~mag~arcsec$^{-2}$ and an estimated half-light radius of 3~kpc, the optical counterpart of AGC~229101 satisfies the selection criteria for a UDG laid out in the \citet{vandokkum15a} paper; in fact its central surface brightness is significantly fainter than those of any of the galaxies in their sample of 47 Coma Cluster UDGs (which range from $\mu_{g,0}$ of $\sim$24$-$26 mag~arcsec$^{-2}$, with a median value of 25.0 mag~arcsec$^{-2}$), and other ALFALFA detected \hi-bearing UDGs (as shown in the right hand panel of Figure \ref{fig:comparison2}). 
The \citet{vandokkum15a} UDGs have absolute magnitudes in the range $M_g$ $=$ $-$12.5 to $-$16.0 mag (median $-$14.3 mag), and stellar masses between 1 x 10$^7$ $\msun$ and 3 x 10$^8$ $\msun$ (median 6 x 10$^7$ $\msun$).  The optical counterpart of AGC~229101 lies within both of these ranges. 

\begin{figure}[t!]
\centering
\includegraphics[width=0.99\columnwidth]{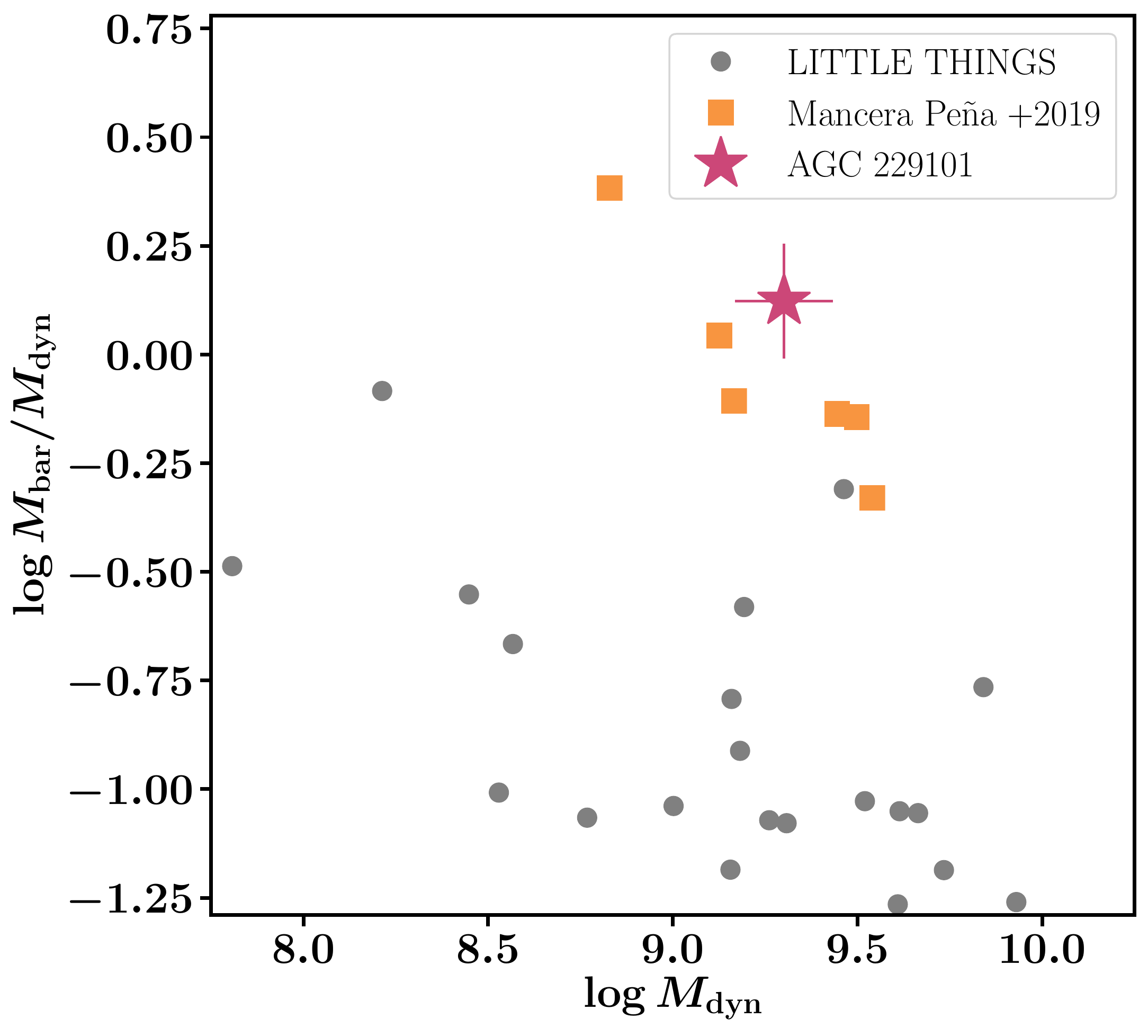}
\caption{Baryon to dynamical mass fraction versus dynamical mass for AGC~229101 (pink star) compared to the ALFALFA UDG sample from \cite{mancerapina19a}, and dwarf galaxies from LITTLE THINGS. While AGC~229101 has a larger gas fraction and lower surface brightness than the \cite{mancerapina19a} sample, it appears to have similarly high baryonic mass for its rotation velocity; within R$_{\rm HI}$ AGC~229101 requires no additional dark matter component.
\label{fig:comparison3}
}
\end{figure}

It seems plausible that hypothesized mechanisms for the formation of galaxies like UDGs, especially ``\hi-bearing UDGs" (low-mass, gas-rich galaxies with low central surface brightnesses and large radii like those described in, e.g., \citealp{dicintio16a}, \citealp{leisman17a}, and \citealp{janowiecki19a}; see Figure \ref{fig:comparison2}) may provide some insight into the origin and evolution of AGC~229101. Importantly, \hi-bearing UDGs have been found to lie off the Baryonic Tully-Fisher relation -- they are rotating too slowly for their baryonic mass \citep{mancerapina19a} -- and to have a wide range of dark matter halo masses, with multiple sources in group environments apparently having little to no detected dark matter. The fact that AGC~229101 has a very narrow velocity profile (43$\pm$9~\kms) despite having 2 x 10$^9$ of \hi\ gas likewise is suggestive of the idea that it has very little dark matter. Indeed, Figure \ref{fig:comparison3} shows a simplistic estimate of the dynamical mass enclosed within the \hi\ radius (see section \ref{sec:discussion:TDG}) places AGC~229101 in a similar parameter space to the \hi-bearing UDGs presented in \cite{mancerapina19a} and \citet{mancerapina20a}, and indicates that AGC~229101 may offer important clues regarding the formation of dark-matter-poor, gas-rich UDGs.

\subsection{Potential Explanations for the Extreme Nature of AGC~229101}
\label{sec:discussion:explanations}

Here we explore a few hypotheses that could help  explain the existence and extreme properties of AGC~229101. 

\subsubsection{AGC~229101: Tidal Dwarf Galaxy?}
\label{sec:discussion:TDG}
Studies have shown that many optically ``dark" \hi\ sources are tidal debris from recent interactions between two galaxies (e.g., \citealp{leismanthesis}). 
Thus one likely explanation for AGC~229101 is that it is a Tidal Dwarf Galaxy (TDG), formed because of 
tidal forces that arose from 
an interaction between a pair of more massive galaxies in its vicinity.
Indeed, considering the masses of the neighboring galaxies around AGC~229101, multiple interaction scenarios are possible. One scenario
is that AGC~229101 could have formed from gas removed from IC~3203 during an interaction with IC~3171. IC~3203 has $10^{10.11}\msun$ of \hi\ \citep{haynes18a}, 
and thus AGC~229101 represents about 15\% of its total gas content, and appears projected on the sky between the two galaxies (see Figure~\ref{fig:environment} and the accompanying discussion in Section~\ref{sec:discussion:environment}).
Another plausible scenario is that AGC~229101 formed as the result of an interaction between IC~3203 and the \hi\ bearing star-forming galaxy IC~3185.

There are multiple pieces of evidence that give weight to the hypothesis that AGC~229101 originated in a tidal encounter between two massive galaxies. 
As mentioned earlier, the low velocity width of AGC~229101 may suggest that the object is dark matter poor. Indeed, if we treat the whole source as if it resides in a single halo and make a naive estimate of the dynamical mass enclosed within its \hi\ radius, we find that $M_{Dyn}(<R_{HI})$ $= R_{HI}~\sigma^2/G$ (where $\sigma = W_{50}/(2\sqrt{2})$) returns a dynamical mass of $\sim$2 x 10$^9 \msun$, which is equal to the enclosed baryonic mass and thus requires no additional dark matter component. 
A commonly used slightly more sophisticated estimate that attempts to account for the velocity distribution and the unknown shape of the total matter distribution is $$M_{dyn} = 2.325 \times10^5 \left(\frac{v_{rot}^2 + 3\sigma^2}{km^2~s^{-2}}\right)\left(\frac{R}{kpc}\right)\msun$$ \citep{hoffman96a}. Assuming $\sigma=11$~\kms, and $v_{rot}=\sqrt{(W_{50}^2-\sigma^2)}/2/\sin(90)$, this gives an estimated dynamical mass of $\sim6.3\times10^9\msun$ within the \hi\ radius. Thus even accounting for the velocity dispersion of the gas, it seems that AGC~229101 is somewhat dark matter poor. 
Moreover, the large and elongated angular extent of the \hi\ cloud also seems reminiscent of tidal tails, and the mass is significantly higher than the typical threshold for forming a long lived TDG \citep{bournaud06a}.  

 
On the other hand, AGC~229101 is at a large separation from IC~3203, without an obvious connecting tail. This large separation to neighboring galaxies suggests that if a tidal hypothesis is correct, AGC~229101 must be a very long-lived tidal system. 
Since IC~3203 is projected at $>$600~kpc on the sky, if we assume typical group velocities of 300 \kms, this would require the neutral gas and stars associated with AGC~229101 to have survived $>$2 Gyr after the galaxy interaction. 


Other TDGs have been detected in clumps at the ends of long tidal tails (e.g., \citet{duc97a}), but not with the large angular extent of the \hi\ clumps in AGC~229101. For example,  
\citet{lee-waddell14a} present the ALFALFA discovery of AGC~749170, an extremely faint TDG. AGC~749290 has a comparable amount of gas to AGC~229101, with a double peaked \hi\ column density distribution and a very low surface brightness, blue optical counterpart. However, in addition to being much closer to its likely progenitor galaxy, AGC~749170 is smaller than AGC~229101 (its half-light radius $\sim$1.5 kpc and its stellar mass is $<10^6\msun$), with the full \hi\ distribution extending $\sim$30~kpc.

The large separation and size of AGC~229101 then may suggest that the southern clump is the remnant of a tidal tail that has somehow managed to survive for an extended period of time. The interaction time scales and distances are made significantly less extreme if one considers the possibility that the northern clump of AGC~229101 is not a TDG, but rather a very low surface brightness source that interacted with IC~3171 several hundred Myr ago. This possibility is interesting, but seems less likely in that the masses in the northern and southern clumps are comparable: an interaction removing the gas in the southern clump from the galaxy would require removing $\sim50\%$ of the galaxy's gas mass. 

\subsubsection{AGC~229101: Other Explanations}
\label{sec:discussion:darks}

While the idea that AGC~229101 is a TDG 
is intriguing, 
there are a number of other potential explanations for the existence of this source. As explained in Section~\ref{sec:discussion:UDG}, the low surface brightness and large radius of AGC~229101 are consistent with, if extreme examples of, the observed properties of \hi\ bearing UDGs (e.g. \citealp{leisman17a,gault21a}). Specifically, the high gas fraction and low rotation velocity are consistent with other reported \hi-bearing UDGs, which appear to lie off the Baryonic Tully Fisher relation (Figure \ref{fig:comparison3}; \citealp{mancerapina19a,mancerapina20a}). Thus proposed mechanisms for the formation of UDGs may apply here.  For example, \citet{amorisco16a} suggest that UDGs are genuine dwarf galaxies that represent the low-mass, high-spin tail of the galaxy distribution.  Alternatively, \citet{dicintio16a} carried out a series of hydrodynamical simulations of galaxy formation and found that UDGs form within moderate mass dark matter halos as a natural consequence of gas outflows that occur along with star formation. 

Another hypothesis that is consistent with our current data is that AGC~229101 could be two merging dark clouds. \citet{starkenburg16a,starkenburg16b} simulate mergers with gas-rich disky dwarf galaxies with dark satellites and follows their evolution to look at star formation, morphology, and kinematics. They find that these simulated systems match the observational properties of many irregular dwarfs and blue compact dwarfs. Though the objects that result from their simulations are more compact than AGC~229101, the data are consistent with the northern clump being a dwarf galaxy interacting with a potential dark cloud (the southern clump). While many of the predicted properties of these systems will require deeper or spectroscopic observations to test, the current observations of AGC~229101 do not rule out this possibility, and even give hints of predicted properties, like misalignment between the optical and \hi\ systems \citep[see][]{starkenburg16b}. 

Finally, we note that we cannot rule out the remote possibility that AGC~229101 is simply an isolated, extended, low surface brightness 
galaxy that resides in a shallow dark matter halo.

%
Deeper high resolution \hi\ imaging of AGC~229101 to study its velocity profile in detail, coupled with deeper optical imaging with next generation telescopes to yield an estimate of the age and metallicity of its stellar component, 
will help to distinguish between the various explanations for the observed properties of this exceptional galaxy. 

\acknowledgments

The authors acknowledge the work of the entire ALFALFA
collaboration in observing, flagging, and extracting sources. 
The authors would also like to thank the anonymous referee for useful comments and suggestions that improved the quality of the manuscript. 
The ALFALFA team at Cornell is supported by NSF grants AST-0607007 
and AST-1107390 to RG and MPH and by grants from the Brinson Foundation. 
The work by HJP, WFJ, NJS, and KLR described in this manuscript was supported by NSF grant AST-1615483 to KLR. SJ acknowledges support from the Australian Research Council’s Discovery Project funding scheme (DP150101734). JMC is supported by NSF grant AST-2009894, and JMC and CB thank Macalester College for support. 
We thank the staff of the WIYN Observatory
and Kitt Peak National Observatory for their help and support during our WIYN pODI observing runs. We are grateful to the staff members at WIYN, NOIRLab, and Indiana University Pervasive Technology Institute for designing and implementing the ODI-PPA and assisting us with the pODI data reduction. Finally, we thank the College of Arts \& Sciences at Indiana University (IU) for providing funding for IU's share of the WIYN telescope. 

This work is based in part on observations made with the VLA, Arecibo Observatory, and WSRT.
The VLA is a facility of the National Radio Astronomy Observatory (NRAO). NRAO is a facility of the National Science Foundation operated under cooperative agreement by Associated Universities, Inc.  The Arecibo Observatory is operated by SRI International under a cooperative agreement with the National Science Foundation (AST-1100968), and in alliance with Ana G. M\`endez-Universidad Metropolitana, and the Universities Space Research Association. 
The Westerbork Synthesis Radio Telescope is operated by the ASTRON (Netherlands Institute for Radio Astronomy) with support from NWO.



%
\bibliography{mybib}

\begin{thebibliography}{}
\expandafter\ifx\csname natexlab\endcsname\relax\def\natexlab#1{#1}\fi
\providecommand{\url}[1]{\href{#1}{#1}}

\bibitem[{{Adams} {et~al.}(2013){Adams}, {Giovanelli}, \& {Haynes}}]{adams13a}
{Adams}, E.~A.~K., {Giovanelli}, R., \& {Haynes}, M.~P. 2013, \apj, 768, 77

\bibitem[{{Amorisco} \& {Loeb}(2016)}]{amorisco16a}
{Amorisco}, N.~C., \& {Loeb}, A. 2016, \mnras, 459, L51

\bibitem[{{Anand} {et~al.}(2018){Anand}, {Tully}, {Karachentsev}, {Makarov},
  {Makarova}, {Rizzi}, \& {Shaya}}]{anand18a}
{Anand}, G.~S., {Tully}, R.~B., {Karachentsev}, I.~D., {et~al.} 2018, \apjl,
  861, L6

\bibitem[{{Ball} {et~al.}(2018){Ball}, {Cannon}, {Leisman}, {Adams}, {Haynes},
  {J{\'o}zsa}, {McQuinn}, {Salzer}, {Brunker}, {Giovanelli}, {Hallenbeck},
  {Janesh}, {Janowiecki}, {Jones}, \& {Rhode}}]{ball18a}
{Ball}, C., {Cannon}, J.~M., {Leisman}, L., {et~al.} 2018, \aj, 155, 65

\bibitem[{{Bell} {et~al.}(2003){Bell}, {McIntosh}, {Katz}, \&
  {Weinberg}}]{bell03a}
{Bell}, E.~F., {McIntosh}, D.~H., {Katz}, N., \& {Weinberg}, M.~D. 2003, \apjs,
  149, 289

\bibitem[{{Bournaud} \& {Duc}(2006)}]{bournaud06a}
{Bournaud}, F., \& {Duc}, P.-A. 2006, \aap, 456, 481

\bibitem[{{Bradford} {et~al.}(2015){Bradford}, {Geha}, \&
  {Blanton}}]{bradford15a}
{Bradford}, J.~D., {Geha}, M.~C., \& {Blanton}, M.~R. 2015, \apj, 809, 146

\bibitem[{{Brinchmann} {et~al.}(2004){Brinchmann}, {Charlot}, {White},
  {Tremonti}, {Kauffmann}, {Heckman}, \& {Brinkmann}}]{brinchmann04a}
{Brinchmann}, J., {Charlot}, S., {White}, S.~D.~M., {et~al.} 2004, \mnras, 351,
  1151

\bibitem[{{Brunker} {et~al.}(2019){Brunker}, {McQuinn}, {Salzer}, {Cannon},
  {Janowiecki}, {Leisman}, {Rhode}, {Adams}, {Ball}, \& {Dolphin}}]{brunker19a}
{Brunker}, S.~W., {McQuinn}, K. B.~W., {Salzer}, J.~J., {et~al.} 2019, \aj,
  157, 76

\bibitem[{{Cannon} {et~al.}(2015){Cannon}, {Martinkus}, {Leisman}, {Haynes},
  {Adams}, {Giovanelli}, {Hallenbeck}, {Janowiecki}, {Jones}, {J{\'o}zsa},
  {Koopmann}, {Nichols}, {Papastergis}, {Rhode}, {Salzer}, \&
  {Troischt}}]{cannon15a}
{Cannon}, J.~M., {Martinkus}, C.~P., {Leisman}, L., {et~al.} 2015, \aj, 149, 72

\bibitem[{{Catinella} {et~al.}(2010){Catinella}, {Schiminovich}, {Kauffmann},
  {Fabello}, {Wang}, {Hummels}, {Lemonias}, {Moran}, {Wu}, {Giovanelli},
  {Haynes}, {Heckman}, {Basu-Zych}, {Blanton}, {Brinchmann}, {Budav{\'a}ri},
  {Gon{\c{c}}alves}, {Johnson}, {Kennicutt}, {Madore}, {Martin}, {Rich},
  {Tacconi}, {Thilker}, {Wild}, \& {Wyder}}]{catinella10a}
{Catinella}, B., {Schiminovich}, D., {Kauffmann}, G., {et~al.} 2010, \mnras,
  403, 683

\bibitem[{{Chengalur} {et~al.}(1995){Chengalur}, {Giovanelli}, \&
  {Haynes}}]{chengalur95a}
{Chengalur}, J.~N., {Giovanelli}, R., \& {Haynes}, M.~P. 1995, \aj, 109, 2415

\bibitem[{{Conselice}(2018)}]{conselice18a}
{Conselice}, C.~J. 2018, Research Notes of the American Astronomical Society,
  2, 43

\bibitem[{{Consolandi} {et~al.}(2016){Consolandi}, {Gavazzi}, {Fumagalli},
  {Dotti}, \& {Fossati}}]{consolandi16a}
{Consolandi}, G., {Gavazzi}, G., {Fumagalli}, M., {Dotti}, M., \& {Fossati}, M.
  2016, \aap, 591, A38

\bibitem[{{Cornwell}(2008)}]{cornwell08a}
{Cornwell}, T.~J. 2008, IEEE Journal of Selected Topics in Signal Processing,
  2, 793

\bibitem[{{Crain} {et~al.}(2017){Crain}, {Bah{\'e}}, {Lagos}, {Rahmati},
  {Schaye}, {McCarthy}, {Marasco}, {Bower}, {Schaller}, {Theuns}, \& {van der
  Hulst}}]{crain17a}
{Crain}, R.~A., {Bah{\'e}}, Y.~M., {Lagos}, C.~d.~P., {et~al.} 2017, \mnras,
  464, 4204

\bibitem[{{Di Cintio} {et~al.}(2017){Di Cintio}, {Brook}, {Dutton},
  {Macci{\`o}}, {Obreja}, \& {Dekel}}]{dicintio16a}
{Di Cintio}, A., {Brook}, C.~B., {Dutton}, A.~A., {et~al.} 2017, \mnras, 466,
  L1

\bibitem[{{Du} {et~al.}(2020){Du}, {Cheng}, {Zheng}, \& {Wu}}]{du20a}
{Du}, W., {Cheng}, C., {Zheng}, Z., \& {Wu}, H. 2020, \aj, 159, 138

\bibitem[{{Du} \& {McGaugh}(2020)}]{du20b}
{Du}, W., \& {McGaugh}, S.~S. 2020, \aj, 160, 122

\bibitem[{{Duc} {et~al.}(1997){Duc}, {Brinks}, {Wink}, \& {Mirabel}}]{duc97a}
{Duc}, P.~A., {Brinks}, E., {Wink}, J.~E., \& {Mirabel}, I.~F. 1997, \aap, 326,
  537

\bibitem[{{Eisenstein} {et~al.}(2011){Eisenstein}, {Weinberg}, {Agol},
  {Aihara}, {Allende Prieto}, {Anderson}, {Arns}, {Aubourg}, {Bailey}, \&
  {Balbinot}}]{eisenstein11a}
{Eisenstein}, D.~J., {Weinberg}, D.~H., {Agol}, E., {et~al.} 2011, \aj, 142, 72

\bibitem[{{Gault} {et~al.}(2021){Gault}, {Leisman}, {Adams}, {Mancera
  Pi{\~n}a}, {Reiter}, {Smith}, {Battipaglia}, {Cannon}, {Fraternali},
  {Haynes}, {McAllan}, {Pagel}, {Rhode}, {Salzer}, \& {Singer}}]{gault21a}
{Gault}, L., {Leisman}, L., {Adams}, E. A.~K., {et~al.} 2021, arXiv e-prints,
  arXiv:2101.01753

\bibitem[{{Giovanelli} \& {Haynes}(1989)}]{giovanelli89a}
{Giovanelli}, R., \& {Haynes}, M.~P. 1989, \apjl, 346, L5

\bibitem[{{Giovanelli} {et~al.}(2010){Giovanelli}, {Haynes}, {Kent}, \&
  {Adams}}]{giovanelli10a}
{Giovanelli}, R., {Haynes}, M.~P., {Kent}, B.~R., \& {Adams}, E.~A.~K. 2010,
  \apjl, 708, L22

\bibitem[{{Giovanelli} {et~al.}(1991){Giovanelli}, {Williams}, \&
  {Haynes}}]{giovanelli91a}
{Giovanelli}, R., {Williams}, J.~P., \& {Haynes}, M.~P. 1991, \aj, 101, 1242

\bibitem[{{Giovanelli} {et~al.}(2005){Giovanelli}, {Haynes}, {Kent},
  {Perillat}, {Saintonge}, {Brosch}, {Catinella}, {Hoffman}, {Stierwalt},
  {Spekkens}, {Lerner}, {Masters}, {Momjian}, {Rosenberg}, {Springob},
  {Boselli}, {Charmandaris}, {Darling}, {Davies}, {Garcia Lambas}, {Gavazzi},
  {Giovanardi}, {Hardy}, {Hunt}, {Iovino}, {Karachentsev}, {Karachentseva},
  {Koopmann}, {Marinoni}, {Minchin}, {Muller}, {Putman}, {Pantoja}, {Salzer},
  {Scodeggio}, {Skillman}, {Solanes}, {Valotto}, {van Driel}, \& {van
  Zee}}]{giovanelli05a}
{Giovanelli}, R., {Haynes}, M.~P., {Kent}, B.~R., {et~al.} 2005, \aj, 130, 2598

\bibitem[{{Gopu} {et~al.}(2014){Gopu}, {Hayashi}, {Young}, {Harbeck},
  {Boroson}, {Liu}, {Kotulla}, {Shaw}, {Henschel}, {Rajagopal}, {Stobie},
  {Knezek}, {Martin}, \& {Archbold}}]{gopu14a}
{Gopu}, A., {Hayashi}, S., {Young}, M.~D., {et~al.} 2014, in \procspie, Vol.
  9152, Software and Cyberinfrastructure for Astronomy III, 91520E

\bibitem[{{Greco} {et~al.}(2018){Greco}, {Goulding}, {Greene}, {Strauss},
  {Huang}, {Kim}, \& {Komiyama}}]{greco18a}
{Greco}, J.~P., {Goulding}, A.~D., {Greene}, J.~E., {et~al.} 2018, \apj, 866,
  112

\bibitem[{{Harbeck} {et~al.}(2014){Harbeck}, {Boroson}, {Lesser}, {Rajagopal},
  {Yeatts}, {Corson}, {Liu}, {Dell'Antonio}, {Kotulla}, {Ouellette}, {Hooper},
  {Smith}, {Bredthauer}, {Martin}, {Muller}, {Knezek}, \&
  {Hunten}}]{harbeck14a}
{Harbeck}, D.~R., {Boroson}, T., {Lesser}, M., {et~al.} 2014, in \procspie,
  Vol. 9147, Ground-based and Airborne Instrumentation for Astronomy V, 91470P

\bibitem[{{Haynes} {et~al.}(2011){Haynes}, {Giovanelli}, {Martin}, {Hess},
  {Saintonge}, {Adams}, {Hallenbeck}, {Hoffman}, {Huang}, {Kent}, {Koopmann},
  {Papastergis}, {Stierwalt}, {Balonek}, {Craig}, {Higdon}, {Kornreich},
  {Miller}, {O'Donoghue}, {Olowin}, {Rosenberg}, {Spekkens}, {Troischt}, \&
  {Wilcots}}]{haynes11a}
{Haynes}, M.~P., {Giovanelli}, R., {Martin}, A.~M., {et~al.} 2011, \aj, 142,
  170

\bibitem[{{Haynes} {et~al.}(2018){Haynes}, {Giovanelli}, {Kent}, {Adams},
  {Balonek}, {Craig}, {Fertig}, {Finn}, {Giovanardi}, {Hallenbeck}, {Hess},
  {Hoffman}, {Huang}, {Jones}, {Koopmann}, {Kornreich}, {Leisman}, {Miller},
  {Moorman}, {O'Connor}, {O'Donoghue}, {Papastergis}, {Troischt}, {Stark}, \&
  {Xiao}}]{haynes18a}
{Haynes}, M.~P., {Giovanelli}, R., {Kent}, B.~R., {et~al.} 2018, \apj, 861, 49

\bibitem[{{Henderson} {et~al.}(1982){Henderson}, {Jackson}, \&
  {Kerr}}]{henderson82a}
{Henderson}, A.~P., {Jackson}, P.~D., \& {Kerr}, F.~J. 1982, \apj, 263, 116

\bibitem[{{Herrmann} {et~al.}(2016){Herrmann}, {Hunter}, {Zhang}, \&
  {Elmegreen}}]{herrmann16a}
{Herrmann}, K.~A., {Hunter}, D.~A., {Zhang}, H.-X., \& {Elmegreen}, B.~G. 2016,
  \aj, 152, 177

\bibitem[{{Hoffman} {et~al.}(1996){Hoffman}, {Salpeter}, {Farhat}, {Roos},
  {Williams}, \& {Helou}}]{hoffman96a}
{Hoffman}, G.~L., {Salpeter}, E.~E., {Farhat}, B., {et~al.} 1996, \apjs, 105,
  269

\bibitem[{{Huang} {et~al.}(2012){Huang}, {Haynes}, {Giovanelli}, \&
  {Brinchmann}}]{huang12b}
{Huang}, S., {Haynes}, M.~P., {Giovanelli}, R., \& {Brinchmann}, J. 2012, \apj,
  756, 113

\bibitem[{{Janesh} {et~al.}(2019){Janesh}, {Rhode}, {Salzer}, {Janowiecki},
  {Adams}, {Haynes}, {Giovanelli}, \& {Cannon}}]{janesh19a}
{Janesh}, W., {Rhode}, K.~L., {Salzer}, J.~J., {et~al.} 2019, \aj, 157, 183

\bibitem[{{Janowiecki} {et~al.}(2019){Janowiecki}, {Jones}, {Leisman}, \&
  {Webb}}]{janowiecki19a}
{Janowiecki}, S., {Jones}, M.~G., {Leisman}, L., \& {Webb}, A. 2019, \mnras,
  490, 566

\bibitem[{{Janowiecki} {et~al.}(2015){Janowiecki}, {Leisman}, {J{\'o}zsa},
  {Salzer}, {Haynes}, {Giovanelli}, {Rhode}, {Cannon}, {Adams}, \&
  {Janesh}}]{janowiecki15a}
{Janowiecki}, S., {Leisman}, L., {J{\'o}zsa}, G., {et~al.} 2015, \apj, 801, 96

\bibitem[{{Jester} {et~al.}(2005){Jester}, {Schneider}, {Richards}, {Green},
  {Schmidt}, {Hall}, {Strauss}, {Vanden Berk}, {Stoughton}, {Gunn},
  {Brinkmann}, {Kent}, {Smith}, {Tucker}, \& {Yanny}}]{jester05a}
{Jester}, S., {Schneider}, D.~P., {Richards}, G.~T., {et~al.} 2005, \aj, 130,
  873

\bibitem[{{Koda} {et~al.}(2015){Koda}, {Yagi}, {Yamanoi}, \&
  {Komiyama}}]{koda15a}
{Koda}, J., {Yagi}, M., {Yamanoi}, H., \& {Komiyama}, Y. 2015, \apjl, 807, L2

\bibitem[{{Kotulla}(2014)}]{kotulla14a}
{Kotulla}, R. 2014, in Astronomical Society of the Pacific Conference Series,
  Vol. 485, Astronomical Data Analysis Software and Systems XXIII, ed.
  N.~{Manset} \& P.~{Forshay}, 375

\bibitem[{{Lee} {et~al.}(2003){Lee}, {McCall}, {Kingsburgh}, {Ross}, \&
  {Stevenson}}]{lee03a}
{Lee}, H., {McCall}, M.~L., {Kingsburgh}, R.~L., {Ross}, R., \& {Stevenson},
  C.~C. 2003, \aj, 125, 146

\bibitem[{{Lee-Waddell} {et~al.}(2014){Lee-Waddell}, {Spekkens}, {Cuillandre},
  {Cannon}, {Haynes}, {Sick}, {Chandra}, {Patra}, {Stierwalt}, \&
  {Giovanelli}}]{lee-waddell14a}
{Lee-Waddell}, K., {Spekkens}, K., {Cuillandre}, J.-C., {et~al.} 2014, \mnras,
  443, 3601

\bibitem[{{Lee-Waddell} {et~al.}(2016){Lee-Waddell}, {Spekkens}, {Chandra},
  {Patra}, {Cuillandre}, {Wang}, {Haynes}, {Cannon}, {Stierwalt}, {Sick}, \&
  {Giovanelli}}]{lee-waddell16a}
{Lee-Waddell}, K., {Spekkens}, K., {Chandra}, P., {et~al.} 2016, \mnras, 460,
  2945

\bibitem[{{Leisman}(2017)}]{leismanthesis}
{Leisman}, L. 2017, PhD thesis, Cornell University.
\newblock \url{https://doi.org/10.7298/X4K935PT}

\bibitem[{{Leisman} {et~al.}(2016){Leisman}, {Haynes}, {Giovanelli},
  {J{\'o}zsa}, {Adams}, \& {Hess}}]{leisman16a}
{Leisman}, L., {Haynes}, M.~P., {Giovanelli}, R., {et~al.} 2016, \mnras, 463,
  1692

\bibitem[{{Leisman} {et~al.}(2017){Leisman}, {Haynes}, {Janowiecki},
  {Hallenbeck}, {J{\'o}zsa}, {Giovanelli}, {Adams}, {Bernal Neira}, {Cannon},
  {Janesh}, {Rhode}, \& {Salzer}}]{leisman17a}
{Leisman}, L., {Haynes}, M.~P., {Janowiecki}, S., {et~al.} 2017, \apj, 842,
  133.
\newblock \url{http://stacks.iop.org/0004-637X/842/i=2/a=133}

\bibitem[{{Mancera Pi{\~n}a} {et~al.}(2019){Mancera Pi{\~n}a}, {Aguerri},
  {Peletier}, {Venhola}, {Trager}, \& {Choque Challapa}}]{mancerapina19a}
{Mancera Pi{\~n}a}, P.~E., {Aguerri}, J.~A.~L., {Peletier}, R.~F., {et~al.}
  2019, \mnras, 485, 1036

\bibitem[{{Mancera Pi{\~n}a} {et~al.}(2020){Mancera Pi{\~n}a}, {Fraternali},
  {Oman}, {Adams}, {Bacchini}, {Marasco}, {Oosterloo}, {Pezzulli}, {Posti},
  {Leisman}, {Cannon}, {di Teodoro}, {Gault}, {Haynes}, {Reiter}, {Rhode},
  {Salzer}, \& {Smith}}]{mancerapina20a}
{Mancera Pi{\~n}a}, P.~E., {Fraternali}, F., {Oman}, K.~A., {et~al.} 2020,
  \mnras, 495, 3636

\bibitem[{{Matsuoka} {et~al.}(2012){Matsuoka}, {Ienaka}, {Oyabu}, {Wada}, \&
  {Takino}}]{matsuoka12a}
{Matsuoka}, Y., {Ienaka}, N., {Oyabu}, S., {Wada}, K., \& {Takino}, S. 2012,
  \aj, 144, 159

\bibitem[{{McMullin} {et~al.}(2007){McMullin}, {Waters}, {Schiebel}, {Young},
  \& {Golap}}]{mcmullin07a}
{McMullin}, J.~P., {Waters}, B., {Schiebel}, D., {Young}, W., \& {Golap}, K.
  2007, in Astronomical Society of the Pacific Conference Series, Vol. 376,
  Astronomical Data Analysis Software and Systems XVI, ed. R.~A. {Shaw},
  F.~{Hill}, \& D.~J. {Bell}, 127

\bibitem[{{Roberts} \& {Haynes}(1994)}]{roberts94a}
{Roberts}, M.~S., \& {Haynes}, M.~P. 1994, \araa, 32, 115

\bibitem[{{Rom{\'a}n} \& {Trujillo}(2017)}]{roman17b}
{Rom{\'a}n}, J., \& {Trujillo}, I. 2017, \mnras, 468, 4039

\bibitem[{{Salim} {et~al.}(2007){Salim}, {Rich}, {Charlot}, {Brinchmann},
  {Johnson}, {Schiminovich}, {Seibert}, {Mallery}, {Heckman}, {Forster},
  {Friedman}, {Martin}, {Morrissey}, {Neff}, {Small}, {Wyder}, {Bianchi},
  {Donas}, {Lee}, {Madore}, {Milliard}, {Szalay}, {Welsh}, \& {Yi}}]{salim07a}
{Salim}, S., {Rich}, R.~M., {Charlot}, S., {et~al.} 2007, \apjs, 173, 267

\bibitem[{{Salzer} {et~al.}(1991){Salzer}, {di Serego Alighieri}, {Matteucci},
  {Giovanelli}, \& {Haynes}}]{salzer91a}
{Salzer}, J.~J., {di Serego Alighieri}, S., {Matteucci}, F., {Giovanelli}, R.,
  \& {Haynes}, M.~P. 1991, \aj, 101, 1258

\bibitem[{{S{\'a}nchez Almeida} {et~al.}(2011){S{\'a}nchez Almeida}, {Aguerri},
  {Mu{\~n}oz-Tu{\~n}{\'o}n}, \& {Huertas-Company}}]{sanchezalmedia11}
{S{\'a}nchez Almeida}, J., {Aguerri}, J.~A.~L., {Mu{\~n}oz-Tu{\~n}{\'o}n}, C.,
  \& {Huertas-Company}, M. 2011, \apj, 735, 125

\bibitem[{{Sault} {et~al.}(1995){Sault}, {Teuben}, \& {Wright}}]{sault95a}
{Sault}, R.~J., {Teuben}, P.~J., \& {Wright}, M.~C.~H. 1995, in Astronomical
  Society of the Pacific Conference Series, Vol.~77, Astronomical Data Analysis
  Software and Systems IV, ed. R.~A. {Shaw}, H.~E. {Payne}, \& J.~J.~E.
  {Hayes}, 433

\bibitem[{{Schlafly} \& {Finkbeiner}(2011)}]{schlafly11a}
{Schlafly}, E.~F., \& {Finkbeiner}, D.~P. 2011, \apj, 737, 103

\bibitem[{{Schlegel} {et~al.}(1998){Schlegel}, {Finkbeiner}, \&
  {Davis}}]{schlegel98a}
{Schlegel}, D., {Finkbeiner}, D., \& {Davis}, M. 1998, in Wide Field Surveys in
  Cosmology, ed. S.~{Colombi}, Y.~{Mellier}, \& B.~{Raban}, 297

\bibitem[{{Serra} {et~al.}(2012){Serra}, {Oosterloo}, {Morganti}, {Alatalo},
  {Blitz}, {Bois}, {Bournaud}, {Bureau}, {Cappellari}, {Crocker}, {Davies},
  {Davis}, {de Zeeuw}, {Duc}, {Emsellem}, {Khochfar}, {Krajnovi{\'c}},
  {Kuntschner}, {Lablanche}, {McDermid}, {Naab}, {Sarzi}, {Scott}, {Trager},
  {Weijmans}, \& {Young}}]{serra12a}
{Serra}, P., {Oosterloo}, T., {Morganti}, R., {et~al.} 2012, \mnras, 422, 1835

\bibitem[{{Starkenburg} {et~al.}(2016{\natexlab{a}}){Starkenburg}, {Helmi}, \&
  {Sales}}]{starkenburg16a}
{Starkenburg}, T.~K., {Helmi}, A., \& {Sales}, L.~V. 2016{\natexlab{a}}, \aap,
  587, A24

\bibitem[{{Starkenburg} {et~al.}(2016{\natexlab{b}}){Starkenburg}, {Helmi}, \&
  {Sales}}]{starkenburg16b}
---. 2016{\natexlab{b}}, \aap, 595, A56

\bibitem[{{Stil} \& {Israel}(2002)}]{stil02a}
{Stil}, J.~M., \& {Israel}, F.~P. 2002, \aap, 389, 29

\bibitem[{{Suess} {et~al.}(2016){Suess}, {Darling}, {Haynes}, \&
  {Giovanelli}}]{suess16a}
{Suess}, K.~A., {Darling}, J., {Haynes}, M.~P., \& {Giovanelli}, R. 2016,
  \mnras, 459, 220

\bibitem[{{Taylor} \& {Webster}(2005)}]{taylor05a}
{Taylor}, E.~N., \& {Webster}, R.~L. 2005, \apj, 634, 1067

\bibitem[{{van Dokkum} {et~al.}(2015){van Dokkum}, {Abraham}, {Merritt},
  {Zhang}, {Geha}, \& {Conroy}}]{vandokkum15a}
{van Dokkum}, P.~G., {Abraham}, R., {Merritt}, A., {et~al.} 2015, \apjl, 798,
  L45

\bibitem[{{Verde} {et~al.}(2002){Verde}, {Oh}, \& {Jimenez}}]{verde02a}
{Verde}, L., {Oh}, S.~P., \& {Jimenez}, R. 2002, \mnras, 336, 541

\bibitem[{{Wang} {et~al.}(2013){Wang}, {Kauffmann}, {J{\'o}zsa}, {Serra}, {van
  der Hulst}, {Bigiel}, {Brinchmann}, {Verheijen}, {Oosterloo}, {Wang}, {Li},
  {den Heijer}, \& {Kerp}}]{wang13a}
{Wang}, J., {Kauffmann}, G., {J{\'o}zsa}, G.~I.~G., {et~al.} 2013, \mnras, 433,
  270

\bibitem[{{Willmer}(2018)}]{willmer18a}
{Willmer}, C. N.~A. 2018, \apjs, 236, 47

\end{thebibliography}
\bibliographystyle{aasjournal}


\end{document}